\documentclass[%
 showkeys,
 10pt,
nofootinbib,
 amsmath,amssymb,
 aps,
pre,
]{revtex4-2}

\usepackage{amsmath}
\usepackage{amsfonts}
\usepackage{amssymb}
\usepackage{placeins}
\usepackage{xcolor}
\usepackage{lineno}
\usepackage{breakcites}
\usepackage{url}

\DeclareMathOperator*{\argmaxB}{\arg\max} % rbp

\usepackage{graphicx}% Include figure files
\usepackage{dcolumn}% Align table columns on decimal point
\usepackage{bm}% bold math
\usepackage[%showframe,%Uncomment any one of the following lines to test 
margin=1.5in,
]{geometry}

\usepackage[hidelinks]{hyperref}

\begin{document}

%\preprint{APS/123-QED}

\title{Bayan Algorithm: Detecting Communities in Networks Through Exact and Approximate Optimization of Modularity} 

\author{Samin Aref}
\altaffiliation[]{Corresponding author's email: aref@mie.utoronto.ca}%
 \affiliation{Department of Mechanical and Industrial Engineering\\ University of Toronto, Toronto, Canada}
\author{Mahdi Mostajabdaveh}%
\affiliation{Department of Mathematical and Industrial Engineering\\ Polytechnique Montr\'eal, Montreal, Canada}%
\author{Hriday Chheda}%
 \affiliation{Department of Mechanical and Industrial Engineering,\\ University of Toronto, Toronto, Canada}%

\begin{abstract}
Community detection is a classic network problem with extensive applications in various fields. Its most common method is using modularity maximization heuristics which rarely return an optimal partition or anything similar. Partitions with globally optimal modularity are difficult to compute, and therefore have been underexplored. Using structurally diverse networks, we compare 30 community detection methods including our proposed algorithm that offers optimality and approximation guarantees: the \textit{Bayan} algorithm. Unlike existing methods, Bayan globally maximizes modularity or approximates it within a factor. Our results show the distinctive accuracy and stability of maximum-modularity partitions in retrieving planted partitions at rates higher than most alternatives for a wide range of parameter settings in two standard benchmarks. Compared to the partitions from 29 other algorithms, maximum-modularity partitions have the best medians for description length, coverage, performance, average conductance, and well clusteredness. These advantages come at the cost of additional computations which Bayan makes possible for small networks (networks that have up to 3000 edges in their largest connected component). Bayan is several times faster than using open-source and commercial solvers for modularity maximization, making it capable of finding optimal partitions for instances that cannot be optimized by any other existing method. Our results point to a few well performing algorithms, among which Bayan stands out as the most reliable method for small networks. A Python implementation of the Bayan algorithm (\textit{bayanpy}) is publicly available through the package installer for Python.\\ %(pip)

This is a post-peer-review accepted manuscript from the journal Physical Review E. The publisher authenticated version (version of record) and full citation details are available on the website of the American Physical Society. \href{https://doi.org/10.1103/PhysRevE.110.044315}{doi.org/10.1103/PhysRevE.110.044315}

\end{abstract}

%Use showkeys class option if keyword

\maketitle

\clearpage

\tableofcontents

\clearpage

\section{Introduction}
\label{s:intro}
Community detection (CD), the data-driven process of partitioning nodes within a network \cite{schaub2017many}, is a core problem in several fields including physics, mathematics, computer science, biology, and other computational sciences \cite{fortunato2022newman}. Among common approaches for CD are the algorithms which are designed to maximize a utility function, \textit{modularity} \cite{newman_modularity_2006}, across all possible ways that the nodes of the input network can be partitioned into an unspecified number of communities. Modularity measures the fraction of edges within communities minus the expected fraction under a random degree-preserving distribution of the edges. Despite their name and design philosophy, current modularity maximization algorithms, which are used by no less than tens of thousands of peer-reviewed studies \cite{Kosowski2020}, generally fail to maximize modularity or guarantee proximity to a globally maximum-modularity partition (an optimal partition) \cite{aref2023suboptimality}. They have a high risk of failing to obtain relevant communities \cite{kawamoto2019counting} and may result in degenerate partitions \cite{good_performance_2010} that are provably far from the underlying community structure \cite{dinh_network_2015,newman_equivalence_2016}. 

Modularity is among the first objective functions proposed for optimization-based CD \cite{newman_modularity_2006,fortunato2016}. Several limitations of modularity \cite{peixoto_2023}, including its resolution limit \cite{fortunato_2007}, have led researchers develop alternative methods for detecting communities using information theory \cite{rosvall_2007,rosvall_2008}, stochastic block modeling (inferential methods) \cite{Karrer_2011,sbm_2014,bpp2020,liu2021scalable,serrano2021community}, and alternative objective functions \cite{surprise_2015,sato_enhanced_2019,marchese2022detecting,paul_community_2022,liu_community_2023}. In spite of its shortcomings, modularity is the most commonly used method for CD \cite{sobolevsky2014general,fortunato2022newman}, while its mathematically rigorous maximization has remained relatively underexplored \cite{aloise_column_2010,dinh_toward_2015,sobolevsky_optimality_2017,aref2023suboptimality}.

We propose the Bayan algorithm: an exact and approximation algorithm for the global maximization of modularity in networks with up to 3000 edges. Bayan resolves a fundamental limitation \cite{aref2023suboptimality} of previous modularity-based community detection algorithms which rely on heuristic rules rather than rigorous mathematical optimization (e.g.\ Integer Programming - IP). We deploy standard techniques from integer programming (and operations research) for solving a network science (and physics) problem. We present in this article the fundamentals of our method using accessible language for network scientists and other practitioners who may deal with optimization for network clustering. We provide detailed references to the supplemental material document for additional technical optimization details.

\section{Exact and approximate vs.\ heuristic modularity maximization}

Maximizing modularity is an NP-hard problem \cite{brandes2007modularity}, which explains the abundance of heuristic algorithms for modularity-based CD. These heuristic algorithms are widely used not only because of their scalability to large networks \cite{zhao2021community}, but also because their high risk of failing to obtain relevant communities is not well understood \cite{kawamoto2019counting}. The scalability of these heuristics come at a cost: their partitions have no guarantee of proximity to an optimal partition \cite{good_performance_2010}. In an earlier study \cite{aref2023suboptimality}, we compared eight modularity-based methods according to the extent to which they maximize modularity. This included the heuristics known as greedy \cite{clauset_finding_2004}, Louvain \cite{blondel_fast_2008}, Leicht-Newman (LN) (a.k.a Reichardt-Bornholdt)%(Reichardt Bornholdt with the configuration model as the null model) 
\cite{rb_pots_2008}, Combo \cite{sobolevsky2014general}, Belief \cite{zhang2014}, Paris \cite{paris_2018}, Leiden \cite{traag_louvain_2019}, and EdMot \cite{edmot_2019}. The results showed that these heuristics rarely return an optimal partition \cite{aref2023suboptimality}. Moreover, their sub-optimal partitions tend to be disproportionally dissimilar to any optimal partition \cite{dinh_network_2015,aref2023suboptimality} according to results based on an \textit{adjusted mutual information} notion of partition similarity \cite{vinh_AMI}. 

There are hundreds of heuristic CD algorithms \cite{fortunato2022newman}, including tens of CD algorithms in publicly available libraries \cite{networkx,graph-tool_2014,rossetti2019cdlib,pygenstability_2023}. However, to the best of our knowledge, there is no study comparing as many as 30 algorithms on standard benchmarks. %This gap often prevents empirical studies from using the most suitable existing method instead of a conveniently accessible or a widely adopted algorithm \cite{Kosowski2020} which is not necessarily suitable for the specific CD task at hand. 
This gap often leads empirical studies to rely on algorithms that are conveniently accessible or widely adopted \cite{Kosowski2020}, rather than selecting the most suitable algorithm for the specific CD task at hand. 

Given the intricacies and nuances of different CD tasks, it is recommended \cite{peel2017ground} to transition from developing general CD algorithms (one-size-fits-all methods) to specialized CD algorithms that perform well on a narrow set of tasks. For the narrow set of small networks with up to 3000 edges (in their largest connected component), advances in IP may push the limits on large networks whose optimal partitions can be obtained using regular computers. While such small networks are sometimes dismissed as having no bearing on practical applications, there are disciplines and application areas where the large majority of networks are small. For example, in psychometric network analysis where community detection is a prevalent method, most networks have fewer than 100 nodes and 89\% of networks have fewer than 30 nodes \cite{christensen2024comparing}. For many offline social networks where the nodes are sentient beings (people, animals, or other social collectives), the data collection methods impose some practical bounds on the size and order of the network. For example, unless archival records are used, it is often not practical to survey or interview more than a few hundred people, especially when the tolerance for missing data is very low. Therefore, we argue that small networks are not just toy examples for fields where networks tend to be large-scale, but they are common examples in several other fields, in which network computations can be made more accurate.

The community detection literature offers much fewer exact and approximation methods\footnote{We make a distinction between heuristic and approximation methods for optimization in that approximation method have guarantees of proximity to optimal solutions.} than heuristics. Each previous exact method has been restricted to a specific network size: Aloise et al.\ have used column generation to find optimal partitions in networks with up to 512 nodes and 819 edges \cite{aloise_column_2010} (within a 27-hour time limit). Dinh and Thai have used linear programming rounding to approximate optimal partitions for networks with 78-2742 edges while reporting that the exact modularity maximization for such networks is cost-prohibitive \cite{dinh_toward_2015}. Sobolevsky et al.\ have reported guaranteed optimal partitions for networks with up to 50 nodes \cite{sobolevsky_optimality_2017}. In 2023, Brusco et al.\ have reported results on optimal partitions for networks with up to 613 edges \cite{brusco2023maximization}. 

\section{Our contributions}

Given the recommended directions \cite{peel2017ground,aref2023suboptimality}, we propose the Bayan algorithm, a specialized CD method capable of providing a globally optimal partition for small networks.

Bayan can alternatively be used to approximate maximum modularity within a user specified factor at a reduced computation time. 
This algorithm is theoretically grounded by an IP formulation of the modularity maximization problem \cite{dinh_toward_2015} and relies on an exact branch-and-cut scheme for solving the NP-hard optimization problem to global optimality. 

We compare Bayan with 29 other CD algorithms.
%on a narrow set of networks within a size limit. 
This also allows us to (1) find other accurate and suitable methods regardless of whether they use modularity or not and (2) assess the practical relevance of modularity as an objective function in comparison to a wide range of other proposed approaches for CD. 

Our proposed algorithm pushes the largest instances whose maximum-modularity partitions can be obtained exactly on an ordinary computer. Bayan handles previously challenging instances of modularity maximization more efficiently. It also solves new instances that cannot be optimized by any other existing methods (see Tables S3--S5 in the SM for the detailed results). %It pushes the limit on the largest instances whose maximum-modularity partitions can be obtained exactly or approximately on an ordinary computer.

\section{Mathematical Preliminaries}

\subsection{Notations}

We represent the simple undirected and unweighted\footnote{For the sake of simplicity, we focus on unweighted graphs in this paper.} %The Bayan algorithm also maximizes modularity in positively weighted graphs.} 
graph $G$ with node set $V$ and edge set $E$ as $G=(V,E)$. Graph $G$ has $|V|=n$ nodes and $|E|=m$ edges. $n$ and $m$ are called the order and size of the graph, respectively. The symmetric adjacency matrix of graph $G$ is represented by $\textbf{A}=[a_{ij}]$ whose entry at row $i$ and column $j$ is $a_{ij}$. Entry $a_{ij}$ indicates whether node $i$ is connected to node $j$ ($a_{ij}=1$) or not ($a_{ij}=0$). The degree of node $i$ is represented by $d_i=\sum_j{a_{ij}}$. The modularity matrix of graph $G$ is represented by $\textbf{B}=[b_{ij}]$ whose entries are $b_{ij} = a_{ij}-{\gamma d_{i}d_{j}}/{2m}$. $\gamma$ is the resolution parameter\footnote{Without loss of generality, we set $\gamma=1$ for all the analysis in this paper. The Bayan algorithm also supports other user-specified $\gamma$ values for multi-resolution modularity maximization.} \cite{lancichinetti_limits_2011}.

Node set $V$ of the input graph $G$ can be partitioned into an unspecified number of $k$ non-overlapping communities\footnote{We focus on the more general CD problem where $k$ is not specified by the user. The desired number of communities can be indirectly controlled by changing $\gamma$ in Bayan.} based on the partition $P=\{V_1,V_2, \dots, V_k \}$ such that $ \bigcup_{1}^k V_i = V$ and $ V_i \cap V_j = \emptyset$. Under partition $P$, the relative community assignment of a pair of arbitrary nodes $(i,j)$, is either same (represented by $1-x_{ij}=\delta(i,j)=1$) or different (represented by $1-x_{ij}=\delta(i,j)=0$). The partition $P$ can therefore be represented as a symmetric partition matrix $\textbf{X}=[x_{ij}]$ with binary entries $x_{ij}$ each indicating the relative community assignment of nodes $i$ and $j$.

\subsection{The optimization problem statement}

Given undirected and unweighted graph $G$ and partition $\textbf{X}$ as input, the modularity function $Q_{(G,\textbf{X})}$ maps its input to a real value in the range of $[-0.5,1]$ according to Eq.\ \eqref{eq0}.

\begin{equation}
\label{eq0}
 Q_{(G,\textbf{X})}= \frac{1}{2m} \sum \limits_{(i,j) \in V^2} \left( a_{ij} - \gamma\frac{d_id_j}{2m}\right) \delta(i,j)
\end{equation}

In the modularity maximization problem for graph $G$, we look for an \textit{optimal} partition: a partition $\textbf{X}^*_{(G)}$ whose modularity is maximum over all possible partitions: $\textbf{X}^*_{(G)}=\argmaxB_{\textbf{X}}Q_{(G,\textbf{X})}$. Any partition of $G$ that in not an optimal partition is a \textit{sub-optimal} partition. 

\subsection{Sparse IP formulation of modularity maximization}

The modularity maximization problem can be formulated as an IP model as in Eq.\ \eqref{eq1} which is proposed in \cite{dinh_toward_2015}. 

\begin{equation}
\label{eq1}
\begin{split}
 &\max_{x_{ij}} Q = \frac{1}{2m} \left( \sum\limits_{(i,j) \in V^2 , i< j} 2b_{ij}(1- x_{ij}) + \sum\limits_{(i,i) \in V^2} b_{ii} \right) \\
&\text{s.t.} \quad x_{ik}+x_{jk} \geq x_{ij} \quad \forall (i,j) \in V^2 , i< j, k\in K(i,j) \\ 
& \quad \quad x_{ij} \in \{0,1\} \quad \forall (i,j) \in V^2 , i< j
\end{split}
\end{equation}

In the IP model in Eq.\ \eqref{eq1} for the input graph $G$, the optimal objective function value $Q^*_{(G)}$ equals the maximum modularity of graph $G$. An optimal partition $\textbf{X}^*_{(G)}$ is represented by the optimal values of the $x_{ij}$ variables. 

% version based on separating sets
In Eq.\ \eqref{eq1}, $K(i,j)$ indicates a minimum-cardinality separating set \cite{dinh_toward_2015} for the nodes $i,j$ and its usage in the IP model of this problem leads to a more efficient formulation with $\mathcal{O}(n^2)$ constraints \cite{dinh_toward_2015} instead of $\mathcal{O}(n^3)$ constraints \cite{brandes2007modularity,agarwal_modularity-maximizing_2008}. 
For every three nodes that satisfy $(i,j) \in V^2 , i< j, k\in K(i,j)$, there are three constraints in the IP model in Eq.\ \eqref{eq1}. These so-called \textit{triangular constraints} collectively ensure that being in the same community is a transitive relation \cite{brandes2007modularity}. 

Solving this optimization problem is NP-hard \cite{brandes2007modularity}, but Bayan uses a branch-and-cut scheme for pushing the limit on the largest instances whose maximum-modularity partitions can be obtained exactly (or approximated within a factor) on an ordinary computer within a reasonable time. A narrative description on how the Bayan algorithm solves the IP model in Eq.\ \eqref{eq1} is provided in Section \ref{s:methodology} while more technical details are provided in the supplemental material (SM) document. A Python implementation of the Bayan algorithm (\textit{bayanpy}) is publicly available through the \href{https://pypi.org/project/bayanpy/}{package installer for Python (pip)}.

\section{The Bayan Algorithm}
\label{s:methodology}

In this section, we provide a narrative description on the main methodological details of the Bayan algorithm. 

\subsection{Triangular constraints as a disjunction}
Given that the decision variables in the IP model in Eq.\ \eqref{eq1} are all binary, the triangular constraints for the triple $(i,j,k)$ can be written as the logical disjunction of Eq.\ \eqref{con:left_cut} and Eq.\ \eqref{con:right_cut}.
\begin{equation}\label{con:left_cut}
 x_{ij} + x_{ik} + x_{jk} = 0
\end{equation}
\begin{equation}\label{con:right_cut}
 x_{ij} + x_{ik} + x_{jk} \geq 2
\end{equation}

\subsection{Branch-and-cut scheme}
To maximize modularity using a branch-and-cut scheme, we start by obtaining lower and upper bounds for the IP model in Eq.\ \eqref{eq1}. Solving the Linear Programming (LP) relaxation resulted from dropping the integrality constraint from the IP model in Eq.\ \eqref{eq1} provides an upperbound. Bayan uses Gurobi \cite{gurobi} to solve all the LP models involved within the branch and cut process. The modularity value of any feasible partition (e.g.\ obtained using the Combo algorithm, given the relative closeness of its partition to optimal partitions \cite{aref2023analyzing}) provides a lower bound. These two values form the bounds for the root node of an IP search tree. 

\subsection{Branching on node triples}
We then select a triple $(i,j,k)$ for which the LP optimal solution violates both Eq.\ \eqref{con:left_cut} and Eq.\ \eqref{con:right_cut} (see Subsection 1.F in the SM for more details on triple selection). On the selected triple, we partition the feasible space into two sub-spaces by adding either Eq.\ \eqref{con:left_cut} or Eq.\ \eqref{con:right_cut} as a cut to the root node problem which leads to a left and a right subproblem. For each subproblem, an upper bound can be obtained by solving the corresponding LP relaxation. Similarly, for each subproblem, a lower bound can be obtained by computing the modularity of a feasible partition obtained using a heuristic (e.g., the Combo algorithm) for the graph modified based on the added cut. 

\subsection{Manipulating the graph for left subproblems}
For the upper bound and lower bound of the left subproblem, we use the analytical results on equivalences in maximum modularity partitions \cite{arenas2007size,lorena2019improving}. Accordingly, we apply Eq.\ \eqref{con:left_cut} by replacing nodes $i,j,k$ with a new supernode and connecting it to all their neighbours. 

\subsection{Manipulating the modularity matrix for right subproblems}
For the lower bound of the right subproblem, we apply Eq.\ \eqref{con:right_cut} by deducting a positive empirically tuned value $\Delta$ from the six modularity matrix entries associated with the triple $(i,j,k)$. This change makes the nodes $i,j,k$ less appealing to be assigned to the same community in the heuristic feasible solution. It does not guarantee that the heuristic solution satisfies Eq.\ \eqref{con:right_cut}, but there is no such requirement for Bayan's convergence to optimality (see Subsection 1.C in the SM for more details on how the $\Delta$ parameter is tuned). %For each subproblem, the actual modularity associated with the feasible partitions of the modified graph provides a lower bound.

\subsection{Upperbound and lowerbound for maximum modularity}
The search tree can further be explored by branching based on another triple whose triangular constraints are violated by the new LP optimal solutions. 
Throughout the search, the largest lower bound is continuously updated and stored as the \textit{incumbent}, while the largest upper bound (of each level) is stored as the \textit{best bound} only after completing the computations for each level of the tree.

\subsection{Closing nodes in the search tree}
Throughout the branching process, three conditions lead to designating a tree node as \textit{fathomed}. First, the LP solution becoming integer. Second, the LP becoming infeasible. Third, the LP solution becoming smaller than the current incumbent. In all these cases, we do not branch on the node and \textit{close} it.

\subsection{Termination of exact and approximate Bayan}
The convergence of incumbent and best bound guarantees that a globally optimal solution has been found. This branch-and-cut scheme gives rise to the \textit{Bayan exact algorithm} if this convergence is used as the termination criterion. Alternatively, users can set a different termination criterion (specifying a desired run time or an optimality gap tolerance) which leads to the \textit{Bayan approximate algorithm} for approximating maximum modularity within an optimality gap (the relative gap between the incumbent and best bound at the termination of the algorithm). 

Section 1 of the SM document provides six detailed subsections on specific technical details about the more nuanced design aspects of the Bayan algorithm. Having provided an accessible explanation of the Bayan algorithm, we move on to describing how standard benchmarks and partition quality measures are used in our study for comparing 30 algorithms including Bayan on an equal footing.

\section{Technical Background for the Comparisons}

\subsection{Partition similarity measure}

We quantify the similarity of a partition obtained by an algorithm to a desirable partition (e.g.\ a planted ground-truth partition) using adjusted mutual information \cite{vinh_AMI} and normalize it symmetrically \cite{jerdee2023normalized}. The normalized adjusted mutual information (AMI) is a measure of similarity of partitions which (unlike normalized mutual information \cite{danon2005comparing,jerdee2023normalized}) adjusts the measurement based on the similarity that two partitions may have by pure chance. AMI for a pair of identical partitions (or permutations of the same partition) equals 1. For two partitions which have no similarity beyond the similarity caused by random chance, AMI takes a small (positive or negative) value close to 0 \cite{vinh_AMI}.

We use AMI because it is shown to be a reliable measure of partition similarity \cite{vinh_AMI,gates2019element}, and avoid using Normalized Mutual Information (NMI) because, despite its common use \cite{roozbahani_community_2023,singh_disintegrating_2022}, several studies indicate that using it leads to incorrect assessments \cite{vinh_AMI,newman2020RMI,gates2019element,mahmoudi2024NMI} and incorrect evaluation of competing algorithms \cite{jerdee2023normalized}. Similar to the NMI, other popularly used measures of partition similarity (the Jaccard index, the Fowlkes-Mallows index, the adjusted Rand index, and the F measure) suffer from at least one form of undesirable bias \cite{gates2019element} and therefore, we avoid using them.

\subsection{Generating structurally diverse benchmark networks}
\label{ss:lfr-abcd}
%400 words are Materials and the following sections 

To evaluate the performance of different community detection algorithms, we use \textit{Lancichinetti-Fortunato-Radicchi} (LFR) benchmark graphs \cite{lancichinetti_benchmark_2008} as well as \textit{Artificial Benchmarks for {Community} {Detection}} (ABCD) graphs \cite{kaminski_artificial_2021}. We generate 1000 synthetic graphs with randomized parameters described below to ensure that the differences observed in performance of the algorithms are not restricted for specific graph structures. 

We generate LFR benchmarks based on the following randomized parameters: number of nodes ($n$) randomly chosen from the range $[20,300]$, maximum degree $\lfloor 0.3n \rfloor$, maximum community size $\lfloor 0.5n \rfloor$, power law exponent for the degree distribution $\tau_1=3$, power law exponent for the community size distribution $\tau_2=1.5$, and average degree of 4. The parameter $\mu$ (mixing parameter) is chosen from the set $\{0.01, 0.1, 0.3, 0.5, 0.7\}$ for five experiment settings (each with 100 LFR graphs). 

ABCD benchmarks \cite{kaminski_artificial_2021} are the more recent alternative to the LFR model for generating benchmarks \cite{lancichinetti_benchmark_2008}. Being directly comparable to LFR, ABCD offers additional benefits including higher scalability and better control for adjusting an analogous mixing parameter \cite{kaminski_artificial_2021}. We generate ABCD benchmarks based on the following randomized parameters: number of nodes ($n$) randomly chosen from the range $[10, 1000)$; minimum degree $d_{min}$ and minimum community size $k_{min}$ randomly chosen from the range $[1, n/4)$; maximum community size chosen randomly from $[k_{min} + 1, n)$; maximum degree chosen randomly from $[d_{min} + 1, n)$; and power law exponents for the degree distribution and community size distribution randomly from $(1, 8)$ and then rounded off to 2 decimal places. The parameter $\xi$ (mixing parameter) is chosen from the set $\{0.1, 0.3, 0.5, 0.7, 0.9\}$ for five experiment settings (each with 100 ABCD graphs). The mixing parameters $\mu$ and $\xi$ can be considered as indicators of the noise in the LFR and ABCD data generation process respectively. 

\subsection{Partition quality measures}
\label{ss:partition-quality}
Retrieval tests based on LFR and ABCD benchmarks provide a level playing field for the community detection algorithms to be compared against each other in a method-agnostic way \cite{lancichinetti_benchmark_2008,kaminski_artificial_2021}. We also provide six partition quality measures for additional comparisons of the 30 CD algorithms without relying on the retrieval of planted partitions.

Comparing CD algorithms based on a single quality function that is used in some of them (e.g.\ description length or modularity) is suggested to be unfair and not rigorous \cite{kaminski_artificial_2021}. Such an approach favors algorithms that were designed based on that specific measure and disfavors other CD algorithms. Given the disagreements on the suitability of partition quality measures \cite{Miasnikof2020}, we refrain from relying on a single measure and instead report all the following measures: description length \cite{bpp2020}, modularity, average conductance \cite{kannan2004clusterings}, coverage \cite{fortunato2010community}, performance \cite{fortunato2010community}, and well clusteredness \cite{Miasnikof2020}.

All these six partition quality measures are functions that take input graph $G$ and input partition $P$ and return a value as the output. For description length and average conductance, a lower value indicates a better partition. For the other four measures, a higher value indicates a better partition. Except description length and modularity, the other four partition quality measures return values in the unit interval. Well clusteredness returns a binary value. We briefly describe these six partition quality measures:

\textbf{Description length}: The \textit{description length} of a dataset refers to the amount of information required to describe the data. According to the minimum description length principle, a model that can compress the data more effectively has captured more of its regularities and is therefore considered to have shown a better performance \citep{hansen2001model}.

To evaluate a partition $P$ of a graph $G = (V, E)$ using a Stochastic Block Model (SBM) \cite{sbm_2014} with parameters $\theta$, the description length is calculated as:
$$
- \ln p(G | \theta, P) - \ln p(\theta, P)
$$
where, $-\ln p(G | \theta, P)$ is the negative log-likelihood of the graph $G$ given the partition $P$ and the SBM parameters $\theta$,
%, representing the amount of information required to describe the edges between communities.
 and $-\ln p(\theta, P)$ is the prior probability of the partition $P$ and the SBM parameters $\theta$. %, representing the amount of information required to describe the partition structure and the model parameters.
In our evaluations, we calculate the description length using the planted partition model introduced in \cite{bpp2020} as the underlying SBM.

\textbf{Modularity}:
The modularity value $Q_{(G,\textbf{X})}$ for graph $G$ and partition $P$ (that corresponds to the partition matrix $\textbf{X}$) is calculated based on Eq.\ \eqref{eq0}.

\textbf{Average conductance}:
The average conductance of partition $P$ and graph $G$ is the mean of conductance values over communities of partition $P$. A value for conductance can be calculated for each community, $i$, of partition $P=\{V_1,V_2, \dots, V_k\}$ and graph $G$ based on $$\frac{|E_i|}{\min (d(V_i),d(V \setminus V_i))}$$ where $|E_i|$ denotes the number of edges with one endpoint in community $V_i$, $d(V_i)$ is the sum of degrees of nodes assigned to community $V_i$, and $d(V \setminus V_i)$ is the sum of degrees of nodes not assigned to community $V_i$. 

\textbf{Coverage}:
The coverage of partition $P$ is the ratio of the number of its intra-community edges to the total number of edges in graph $G$.

\textbf{Performance}:
The performance of partition $P$ is the total count of intra-community edges plus inter-community non-edges divided by $n(n-1)/2$. This denominator is the size of the complete graph that has $n$ nodes just like $G$.

\textbf{Well clusteredness}:
The \textit{well clusteredness} is a binary measure based on the the three measures $\bar{K}_\text{inter}$, $K$, and $\bar{K}_\text{intra}$ proposed in \cite{Miasnikof2020} and defined as follows:

$$\bar{K}_\text{inter}(G,P)=\frac{2}{k(k-1)}\sum_{i=1}^k\sum_{j=i+1}^k \frac{|E_{ij}|}{|V_i||V_j|}$$ 

where $k$ is the number of communities in partition $P$, $V_i$ denotes the i-th community in partition $P$, $|E_{ij}|$ denotes the number of edges with one endpoint in community $i$ and one endpoint in community $j$;

$$K(G)={2m}/{n(n-1)}$$
where $K(G)$ is the density of graph $G$; and

$$\bar{K}_\text{intra}(G,P)=\frac{\sum_{i=1}^k K(V_i)}{k}$$
where $K(V_i)$ denotes the density of community $V_i$.

The necessary condition for graph $G$ to be \textit{well clustered} by partition $P$ is for the two inequalities $\bar{K}_\text{inter} < K < \bar{K}_\text{intra}$ to hold \cite{Miasnikof2020}. For any graph $G$ and partition $P$, the binary measure, well clusteredness, takes 1 if graph $G$ is well clustered by partition $P$, and it takes 0 otherwise.

\subsection{Comparing CD algorithms}
There are different views on how CD algorithms can be assessed. The two major, and often opposing views, rely on theoretical argumentation vs. practical argumentation.

Some experts use a theoretical argumentation and suggest that when we generate networks using some generative process, like the LFR or ABCD benchmark, then provably the Bayes-optimal algorithm for inferring the planted communities is the one that simply inverts the generative process for inference. Therefore, without performing any comparisons, we know that Maximum A Posteriori estimation with a Degree-Corrected SBM with appropriate priors and mixing structure will perform the best for the LFR benchmark if an infinite number of trials are conducted and we use a suitable error measure for the performance (an ideal partition similarity measure). They argue that similar guarantees hold for other estimators and loss functions. This argument concludes that any distinction among algorithm performance is thus a finite sample size issue or because the error measure used is different (e.g. AMI or NMI). 

Other experts use a practical argumentation and suggest that standard LFR and ABCD benchmarks are informative for comparing the capabilities of CD algorithms \cite{lancichinetti_benchmark_2008,kaminski_artificial_2021}. LFR and ABCD are standard and widely adopted benchmarks developed to serve this specific purpose of comparing the practical capabilities of CD algorithms in a controlled environment. %The degree-corrected SBM being the Bayes-optimal algorithm under stringent conditions does not solve the community detection problem. 
In practical situations, appropriate priors and mixing structure cannot be simply assumed and an infinite number of trials cannot be performed. Therefore, the LFR and ABCD benchmarks provide valuable insights on the performance of CD algorithms that may violate theoretical expectations. 
Moreover, if we generate benchmarks and use the information on how they are generated (LFR/ABCD) to reach analytical results for supporting a specific algorithm, we would defeat the purpose of the benchmark models. 

Our problem definition in Section \ref{s:intro} is aligned with the practical perspective rather than the theoretical perspective. Therefore, after acknowledging the merits and limitations of the theoretical perspective that suggests comparing CD algorithms is fruitless, we continue to compare 30 CD algorithms using standard LFR and ABCD benchmarks as well as six partition quality measures to assess their practical capabilities on structurally diverse benchmarks across different mixing parameter values. We briefly describe the motivations behind the development of several benchmark models for community detection to prepare the context for the interpretation of our results in Section \ref{s:retrieval}. 

The LFR benchmarks are developed with a timely observation on the part of its developers arguing that ``many algorithms have been proposed [$\dots$] the question of how good an algorithm is, with respect to others, is still open'' \cite[pp 046110-1]{lancichinetti_benchmark_2008}. The LFR benchmark model attempt to address a methodological problem of crucial importance \cite{staudt2017generating,Slota2019}: comparing the accuracy of CD methods. The LFR benchmark improved upon earlier benchmarks  \cite{girvan-newman-2002} which were preferential attachment networks of 128 nodes with fixed degrees and fixed communities sizes. 

The development of the ABCD benchmarks was motivated by the necessity of comparing CD algorithms on synthetic graphs that resemble some key features of typical real-world networks like community structure, heterogeneous degrees, and heterogeneous community sizes \cite{kaminski_artificial_2021}. Algorithms should be compared based on networks with a varying strength of community structure which is operationalized through the $\xi$ mixing parameter. Emphasizing on the usefulness of the LFR benchmark model, ABCD builds upon the same foundation, but resolves three limitations of the LFR model \cite{kaminski_artificial_2021}. 

%The 30 CD algorithms are described next.

\subsection{CD algorithms included in our evaluations}
Using LFR and ABCD as two different families of benchmarks, six partition quality measures, and four real networks from different contexts, we compare the following 30 CD algorithms: Bayan, eight modularity-based heuristics, other optimization-based algorithms, and CD methods which do not rely on modularity or optimization. The 29 algorithms compared to Bayan (in the chronological order of 1970-2023) are the CD methods known as 
Kernighan-Lin bisection \cite{kernighan1970efficient},
greedy \cite{clauset_finding_2004},
Chinese whispers \cite{chinesewhispers_2006},
Reichardt Bornholdt with Erd\H{o}s-R\'{e}nyi as the null model (RB) \cite{rber_pots_2006}, 
Walktrap \cite{walktrap_2006}, 
k-cut \cite{kcut_2007}, 
asynchronous label propagation \cite{asynchronous_label_propagation2007},
Louvain \cite{blondel_fast_2008}, 
Infomap \cite{rosvall_2008}, 
Reichardt Bornholdt with the configuration model as the null model (LN) \cite{rber_pots_2006,rb_pots_2008}, 
Genetic Algorithm (GA) \cite{ga_2008}, 
semi-synchronous label propagation \cite{label_propagation_2010}, 
CPM \cite{cpm_2011}, 
significant scales \cite{significance_communities_2013}, 
Weighted Community Clustering (WCC) \cite{scd_2014}, 
Combo \cite{sobolevsky2014general}, 
Belief \cite{zhang2014}, 
Stochastic Block Model (SBM) (minimum description length) \cite{sbm_2014},
SBM with Markov Chain Monte Carlo (MCMC) \cite{sbm_2014},
Surprise \cite{surprise_2015}, 
Diffusion Entropy Reducer (DER) \cite{der_2015}, 
Paris \cite{paris_2018}, 
Leiden \cite{traag_louvain_2019}, 
EdMot \cite{edmot_2019}, 
GemSec \cite{gemsec_2019},
Bayesian Planted Partition (BPP) \cite{bpp2020},
BPP with MCMC \cite{bpp2020},
Markov stability (PyGenStability\footnote{The Markov stability algorithm in the PyGenStability library produces multiple partitions at different scales for one input network \cite{pygenstability_2023}. We use two simple adaptations of it (MR and MV) to obtain one partition for each input network. These two adaptations involve running PyGenStability with its automated optimal scale selection \cite{pygenstability_2023} which returns multiple candidate partitions corresponding to different scales. In MV, we select the partition with the minimum normalized variance of information. In MR, we select a partition randomly.}) \cite{pygenstability_2023} with random partition selection (MR), and Markov stability \cite{pygenstability_2023} with minimum normalized variance of information (MV). 

We use the public implementations of these 29 algorithms from the following Python libraries. For the two Markov stability algorithms (MR and MV), we use the \textit{PyGenStability} library (version 0.2.2) \cite{pygenstability_2023}. For the four inferential methods (SBM, SM, BPP, and BM), we use the \textit{graph\_tool} library (version 2.57) \cite{graph-tool_2014}. For the two algorithms, Kernighan-Lin and asynchronous label propagation, we use the \textit{NetworkX} library (version 3.1) \cite{networkx}. For the remaining 21 algorithm, we use the Community Discovery library (\textit{CDlib}) (version 0.2.6) \cite{rossetti2019cdlib}.

Three comparisons are provided on these 30 algorithms in Sections \ref{s:retrieval} -- \ref{s:time}.

\section{Partition Retrieval Comparisons}\label{s:retrieval}

This section provides the first of the three comparisons which assesses the 30 algorithms on the extent to which they retrieve planted partitions of LFR and ABCD networks. 

\subsection{Comparison of 30 algorithms based on retrieving planted communities in LFR graphs}
\label{ss:results_lfr}

In our first retrieval comparison, we use the LFR benchmarks \cite{lancichinetti_benchmark_2008} introduced and parameterized earlier, and measure the performance of each method using the AMI of the partition with the planted partition from the LFR graph generation process. 

In this evaluation, we use 500 LFR benchmark graphs in five experiment settings; each having 100 LFR graphs generated based on a specific $\mu$ value ($\mu \in\{0.01,0.1,0.3,0.5,0.7\}$). %Additional details on generating LFR benchmarks are provided in the Materials and Methods section. 
The mixing parameter $\mu$ in the LFR model determines the fraction of inter-community edges. Larger values of $\mu$ complicate the accurate discovery of communities by decreasing the association between the structure and the planted communities. %Given this, $\mu$ can be considered as an indicator of the noise in the data generation process.

\begin{figure*}
 \centering
 \includegraphics[width=0.8\textwidth]{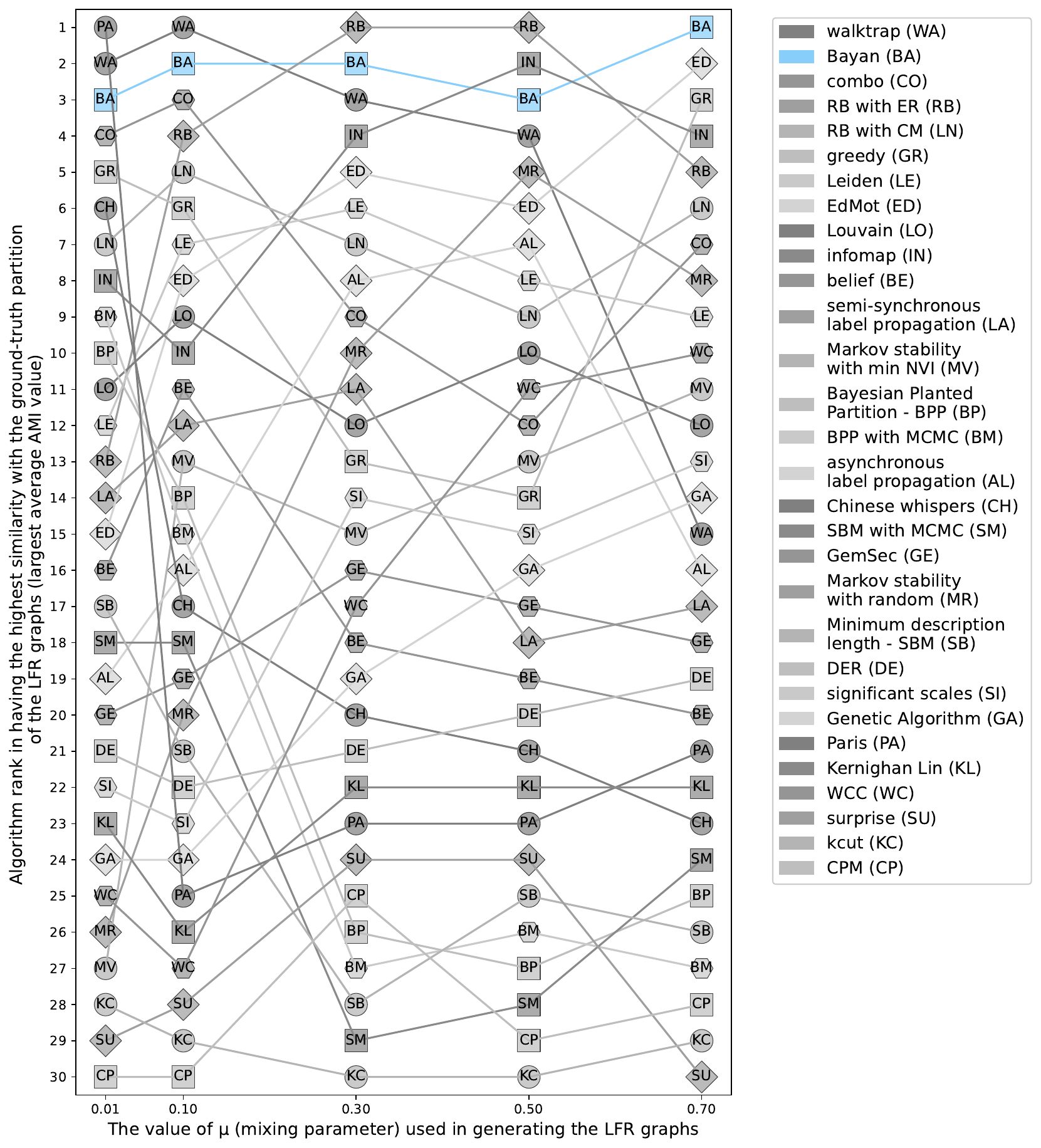}
 \caption{Performance ranking of 30 CD algorithms based on AMI averaged over 100 LFR graphs for each data point. (Magnify the high-resolution figure on screen for details.)}
 \label{fig:RankingLFR}
\end{figure*}

\begin{figure*}
 \centering
 \includegraphics[width=0.8\textwidth]{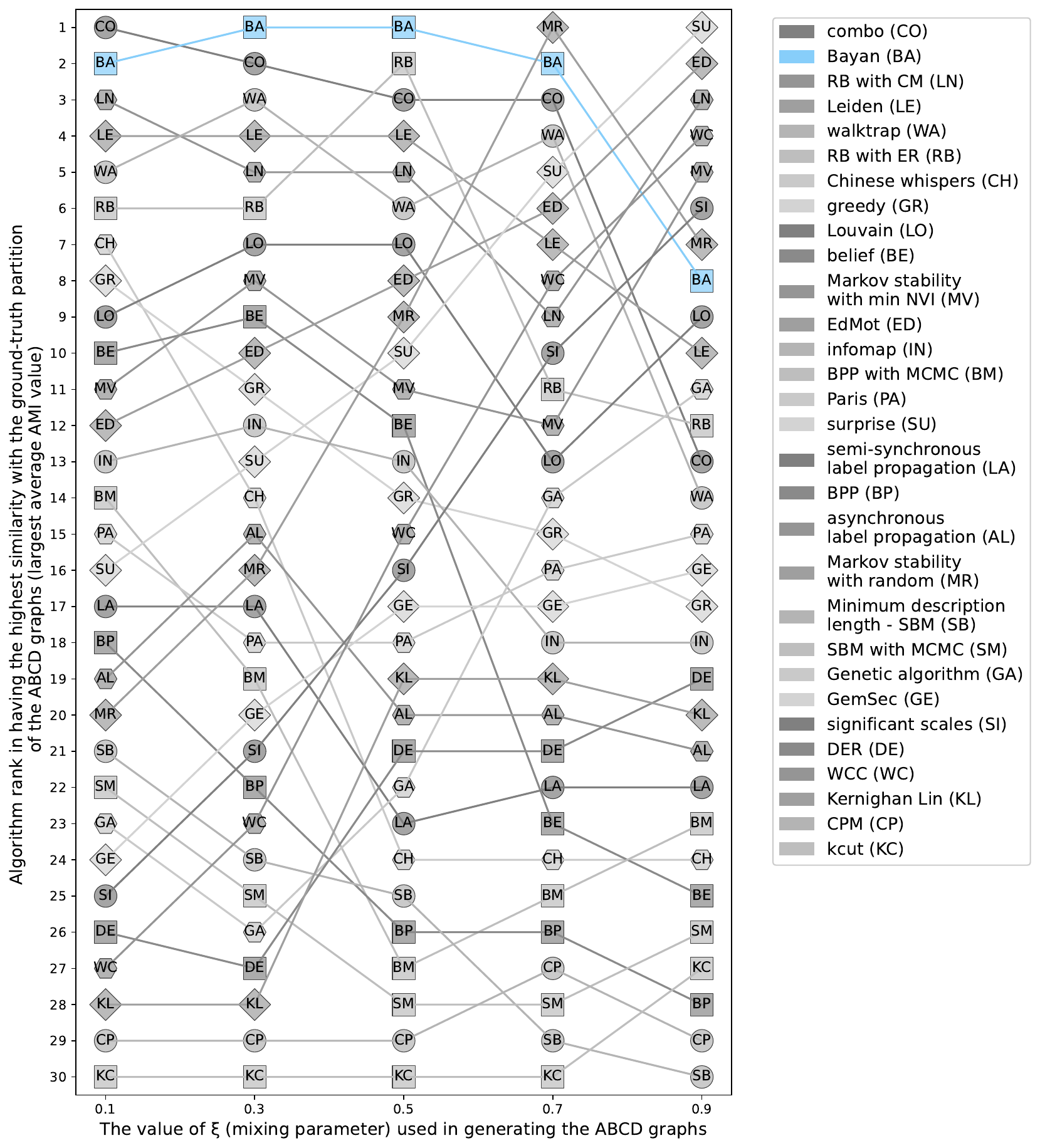}
 \caption{Performance ranking of 30 CD algorithms based on AMI averaged over 100 ABCD graphs for each data point. (Magnify the high-resolution figure on screen for details.)}
 \label{fig:RankingABCD}
\end{figure*}

Results on the average AMI of each algorithm in each experiment setting are provided in the appendix (Table I). Figure~\ref{fig:RankingLFR} illustrates the ranking of the 30 algorithms, including Bayan, according to AMI averaged over 100 LFR graphs in each of the five experiment settings (five values of $\mu$). Among the 30 algorithms, Bayan is always among the three algorithms with the highest average AMIs. Bayan also has the most stable performance across the five values of $\mu$. Walktrap and RB are two other algorithms demonstrating high performance; each of them achieves average AMI higher than Bayan's on two out of five experiment settings. However, their performance ranking across the five values of $\mu$ is less stable than Bayan's.

In Figure~\ref{fig:RankingLFR}, the two algorithms Paris and Combo seem to perform comparatively well for small values of $\mu$, but their performance rankings decrease substantially when the $\mu$ mixing parameter passes the threshold value of $0.1$. Contrary to this pattern, there are algorithms like EdMot whose comparative performance improves when $\mu$ increases. Infomap and three modularity-based heuristics, Combo, LN, and Leiden, show a comparatively decent performance, having higher AMIs than most other methods.
Our observations on the high performance of modularity-based methods on LFR graphs are aligned with the results on LFR graphs reported in \cite{jerdee2023normalized}, where modularity-based methods return AMIs higher than asynchronous label propagation, walktrap, Bayesian planted partition, and sometimes higher than Infomap.

The descriptive AMI ranking results of the 30 algorithms in each experiment setting are validated using a Friedman test of multiple groups \cite{friedman1937} followed by a post-hoc Li test of multiple hypotheses \cite{li2008multiple}. More information about these statistical procedures is provided in Section 2 of the SI. The AMI results on the 100 LFR graphs with $\mu=0.01$ with Bayan as the control method lead to the rejection of the null hypotheses in 19 out of 29 post-hoc Li tests. This suggests that for LFR graphs with $\mu=0.01$, Bayan returns partitions with statistically significant higher similarity to the planted partition compared to each of those 19 other algorithms. For $\mu=0.1$, the null hypotheses are rejected in 28 out of 29 post-hoc Li tests. The only exception is the Walktrap algorithm, for which there is insufficient evidence of difference from Bayan in AMI. The limited rejections are partly due to the limitations in the statistical power of conducting a large number of tests among 30 algorithms while controlling the family-wise error rate to $0.05$ \cite{li2008multiple}. Overall, the statistical results (in Table \ref{tab:1} in the SM) show that Bayan has a significantly higher AMI means in comparison to most of the 29 other algorithms considered confirming the descriptive results in Figure~\ref{fig:RankingLFR} (and Table I in the Appendix).

\subsection{Comparison of 30 algorithms based on retrieving planted communities in ABCD graphs}
\label{ss:results_abcd}

We conduct additional retrieval tests to ensure that the results are not artifacts of using LFR benchmarks. In our second retrieval comparison, we use the ABCD benchmarks \cite{kaminski_artificial_2021} introduced and parameterized earlier, and compare Bayan to the same set of 29 other CD algorithms and rank them based on their average AMI with the planted partitions on ABCD benchmark graphs. 

To evaluate the 30 algorithms based on retrieval of the planted communities, we use 500 ABCD benchmark graphs in five experiment settings; each having 100 ABCD graphs generated based on a specific $\xi$ value ($\xi\in\{0.1,0.3,0.5,0.7,0.9\}$). The mixing parameter $\xi \in [0,1]$ determines the independence of the edge distribution from the communities \cite{kaminski_artificial_2021}. %$\xi$ linearly correlates with the fraction of inter-community edges. 
At the extreme value of $\xi=0$, all edges are within communities. In contrast, at the other extreme value, $\xi=1$ indicates that communities do no influence the distribution of edges \cite{kaminski_artificial_2021}. Therefore, larger values of $\xi$ complicate the discovery of communities by decreasing the association between the structure and the planted communities. %Given this, $\xi$ can be considered as an indicator of the noise in the data generation process.

Results on the average AMI of each algorithm in each experiment setting are provided in the appendix (Table II). Figure~\ref{fig:RankingABCD} illustrates the comparative ranking of the algorithms, including Bayan, according to AMI values averaged over 100 ABCD graphs in each of the five experiment settings (five values of $\xi$). Bayan has the highest or the second highest average AMI in four out of five experiment settings. Among the 30 algorithms for $\xi\in\{0.1,0.3,0.5,0.7\}$, Bayan and Combo are consistently in the top three algorithms with the highest average AMIs. Setting aside the high-noise experiment setting of $\xi=0.9$ where AMIs substantially drop across the board, Bayan and Combo have the most stable performance in accurate retrieval of the planted partitions of these ABCD graphs. 

In Figure~\ref{fig:RankingABCD}, the three algorithms walktrap, LN, and Leiden seem to perform comparatively well especially for small values of $\xi$. Contrary to this pattern, there are algorithms like EdMot and surprise whose comparative performance monotonically improves when $\xi$ increases. Averaging the AMI values for all 500 ABCD graphs, Bayan has the highest total average AMI followed by Combo and Leiden respectively. Note that these three high-performing algorithms are all based on modularity while Bayan differs from the others in having optimality guarantees.

\begin{figure}
 \centering
\includegraphics[width=0.6\textwidth]{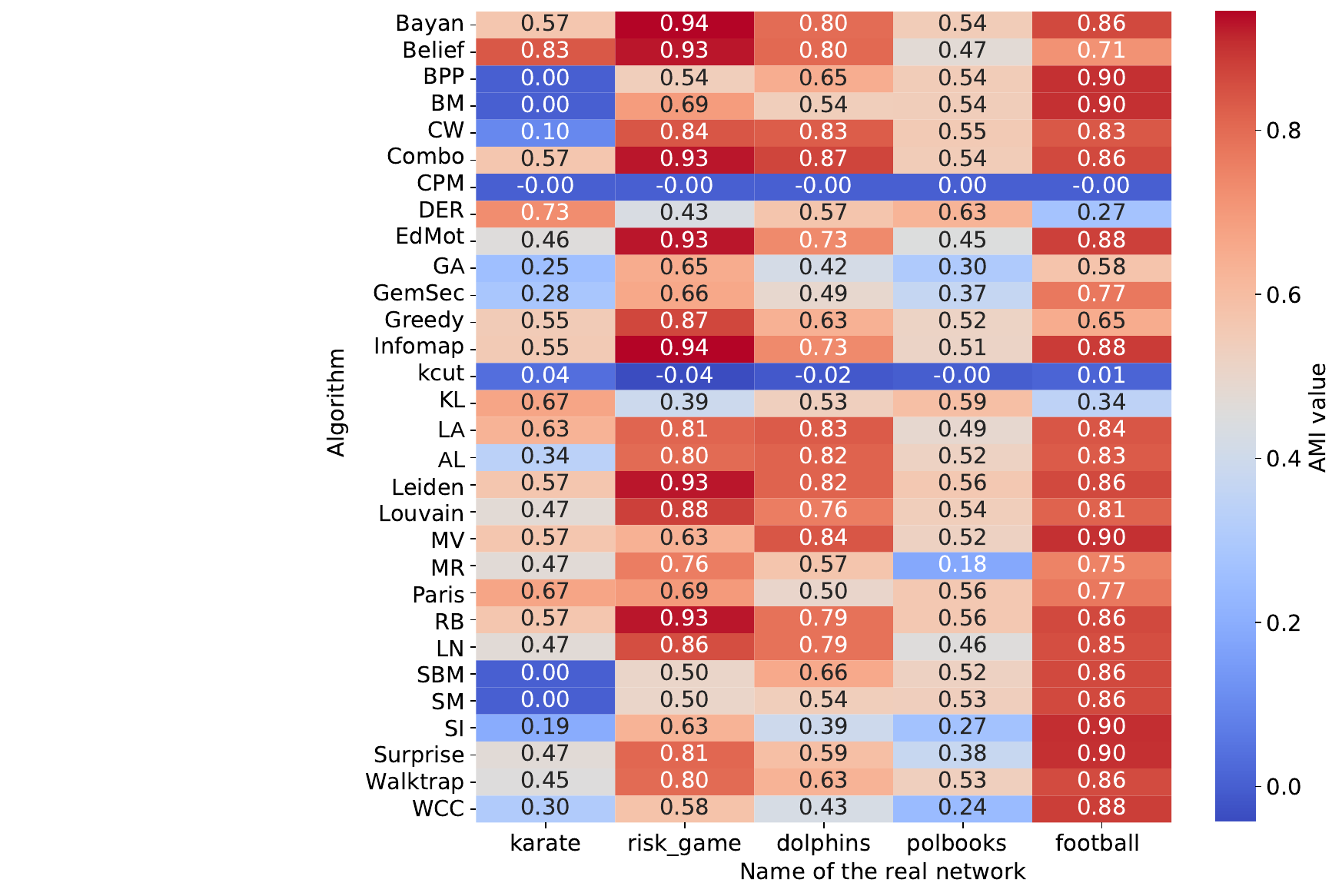}
 \caption{Adjusted mutual information (AMI) values of the CD algorithms indicating the similarity of their partitions with node attributes for five real networks (Color version online. Magnify the high-resolution figure on screen for details.)}
 \label{fig:heatmap}
\end{figure}

Similar to the rankings on LFR graphs, the descriptive ranking results on ABCD graphs are validated using a Friedman test of multiple groups \cite{friedman1937} followed by a post-hoc Li test of multiple hypotheses \cite{li2008multiple}. The AMI results on the 100 ABCD graphs with $\xi=0.1$ with Bayan as the control method lead to the rejection of the null hypotheses in 19 out of 29 post-hoc Li tests. This suggests that for ABCD graphs with $\xi=0.1$, Bayan statistically significantly differs from each of the 19 other algorithms in retrieving the ground truth partition. The exceptions are ten algorithms for which there is insufficient evidence of statistically significant difference from Bayan in AMI. %This is partly due to the limitations in the statistical power of conducting a large number of tests among 30 algorithms while controlling the family-wise error rate to $0.05$ \cite{li2008multiple}. 
Detailed statistical test results on ABCD graphs are provided in the SM (Table \ref{tab:2}) confirming the results in Figure~\ref{fig:RankingABCD} (and Table II in the Appendix).

Take together with the results on LFR graphs, maximum-modularity partitions are shown to have a distinctive accuracy in retrieving planted partitions across a wide range of parameter settings for both standard CD benchmark models. %Under the assumption that LFR and ABCD benchmarks are reasonably informative approaches for comparing algorithms, 
Our results in Figures~\ref{fig:RankingLFR}--\ref{fig:RankingABCD} (and Tables I--II) show the major differences between the extent to which these 30 algorithms retrieve planted partitions in practice where the sample size is not infinite.

Designed as a specialized algorithm for solving an NP-hard problem in small networks (with up to 3000 edges), Bayan is not suitable for and does not scale to large-scale networks. Figures \ref{fig:RankingLFR}--\ref{fig:RankingABCD} also provide insight on algorithms other than Bayan that have reasonably good performance on both LFR and ABCD assessments. At the low extreme of performance ranks, there are several algorithms whose capability of retrieving ground-truth communities is not better than that of the Kernighan-Lin bisection algorithm from 1970. 
Note that for comparing all these 30 algorithms, including several algorithms that do not (and cannot possibly) scale to large instances, we could not include large-scale benchmark networks in our assessments. Given this limitation, our results are safer to be interpreted within the context of the comparative performance of 30 algorithms on small networks with up to 3000 edges. Moreover, from a theoretical standpoint, the finite sample size prevents us from making a firm determination about how these algorithms compare asymptotically.

\subsection{Comparison of 30 algorithms based on similarity to node attributes of fairly modular real networks}
\label{ss:results_real}

We also assess the similarity between the partitions of the same 30 algorithms and a set of node metadata (discrete-valued node attributes) on five real benchmark networks which are commonly used \cite{hric2014community,yang2015defining,newman2016structure,sobolevsky_optimality_2017,kawamoto2018comparative,chen_global_2018,edmot_2019,sobolevsky2022gnn} in the literature for demonstrating the output of CD algorithms. Retrieving node label communities is not the purpose of CD algorithms, and network formation may poorly correlate with any arbitrary chosen node attribute \cite{peel2017ground}. Therefore, we do not consider a given set of node metadata as ground truth. However, it is useful to compare the partitions returned by different algorithms based on their similarity with node metadata \cite{yang2015defining,newman2016structure} to evaluate them in a setting without synthetic benchmarks on networks where node labels are aligned with the structure. Consistent with the literature \cite{edmot_2019,chen_global_2018,hric2014community,steinhaeuser2010identifying}, we use the five real networks commonly referred to as \textit{dolphins}, \textit{football}, \textit{karate}, \textit{polbooks}, and \textit{risk\_game} which all have node attributes. Information on accessing all networks are in Section 5 of the SI. 

Figure~\ref{fig:heatmap} shows the AMI of the partitions from each of the 30 algorithms and the node attributes of the networks. As the node attribute considered is at most one factor among a multitude of factors determining the formation of these networks, the AMI values confound several effects and cannot be reliably interpreted as a performance measure of the algorithms \cite{peel2017ground}. They simultaneously measure the metadata's relevance to the structure and the performance of the algorithms in retrieving communities corresponding to the structure. Bearing these limitations in mind, the results indicate that some algorithms including Bayan return partitions with reasonably high similarity to the node attributes for all five networks, whereas, other algorithms like CPM, k-cut, and SBM return partitions with a less straightforward correspondence to the node metadata. Some algorithms like walktrap, which comparatively had good performance on LFR and ABCD graphs, return some partitions with low similarity to node attributes in this assessment. A No-Free-Lunch theorem for CD \cite{peel2017ground} implies that no single method can consistently outperform others in accurately retrieving the possible node attributes of real networks. Maximum-modularity partitions are (happen to be) reasonably similar to the node attributes for these networks.

So far, we have discussed the advantage of maximum-modularity partitions in retrieving planted communities in LFR and ABCD graphs, and discovering partitions with strong associations with node attributes in some commonly used real networks. Next, we report six comparisons of the 30 algorithms based on partition quality measures which do not depend on partition similarity measures. 

\section{Partition Quality Comparisons}\label{s:quality}

This section provides the second comparison out of the three comparisons of the 30 CD algorithms. Here, we assess the algorithms based on six partition quality measures on the same LFR and ABCD networks. 

\begin{figure}
    \centering
    \includegraphics[width=0.8\linewidth]{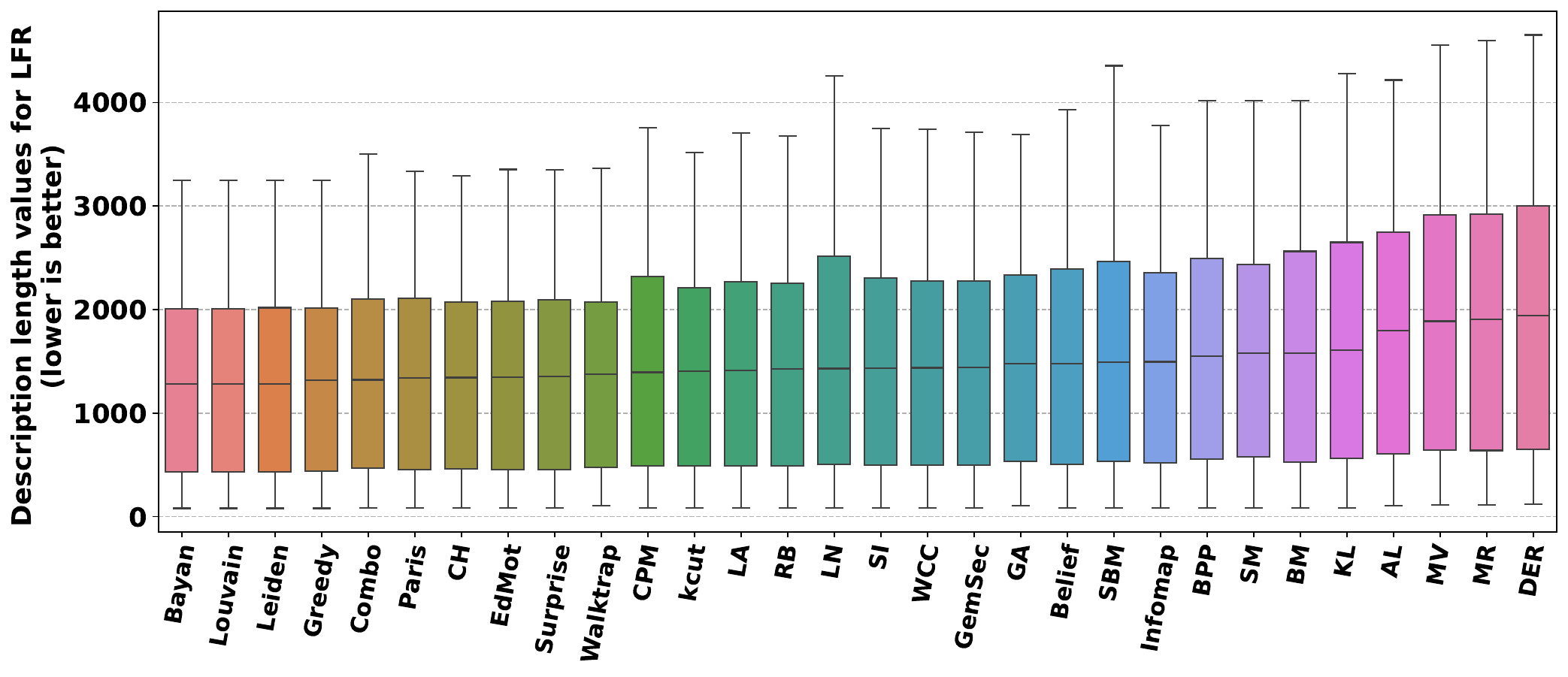}
    \includegraphics[width=0.8\linewidth]{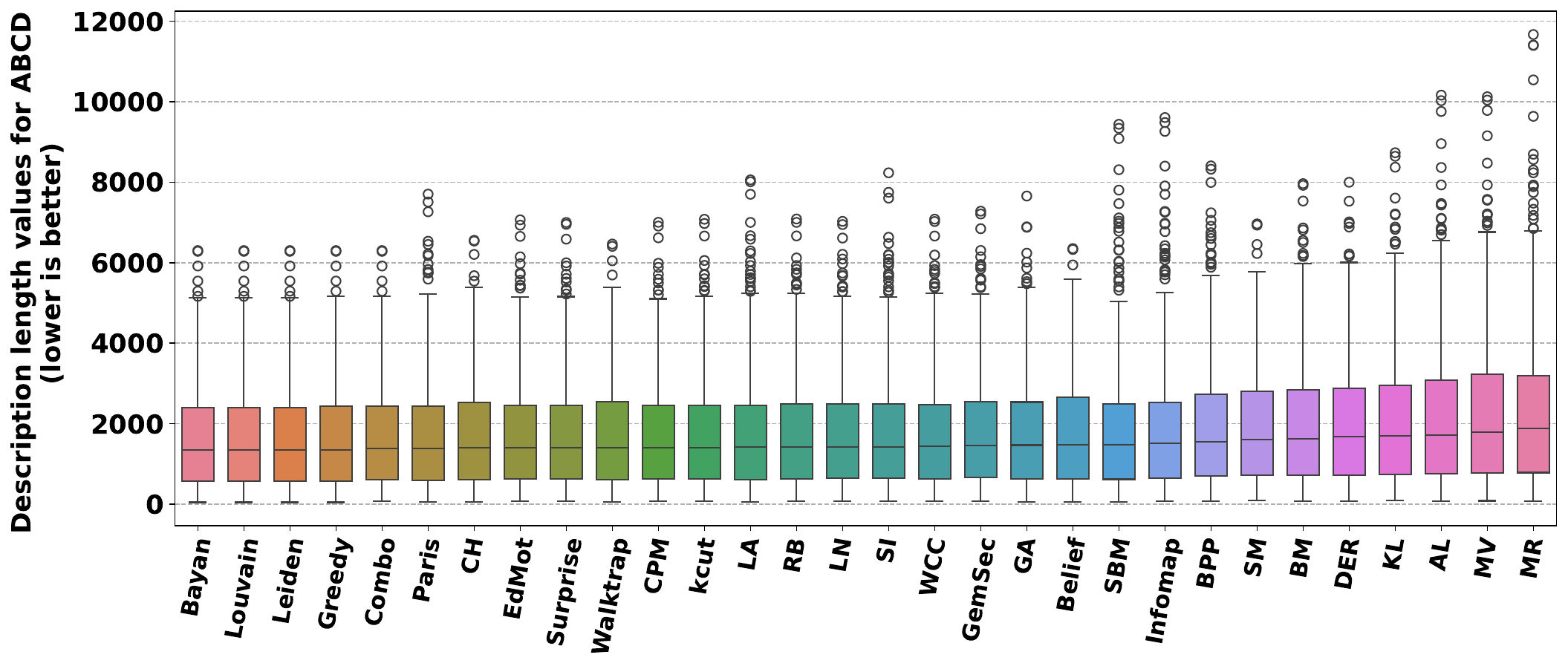}
    \caption{The distribution of description length values for the partitions produced by each algorithm on the 500 LFR networks (top panel) ad the 500 ABCD networks (bottom panel). The algorithms are sorted from right to left based on having a more desirable median description length.}
    \label{fig:description-length}
\end{figure}

\begin{figure}
    \centering
    \includegraphics[width=0.8\linewidth]{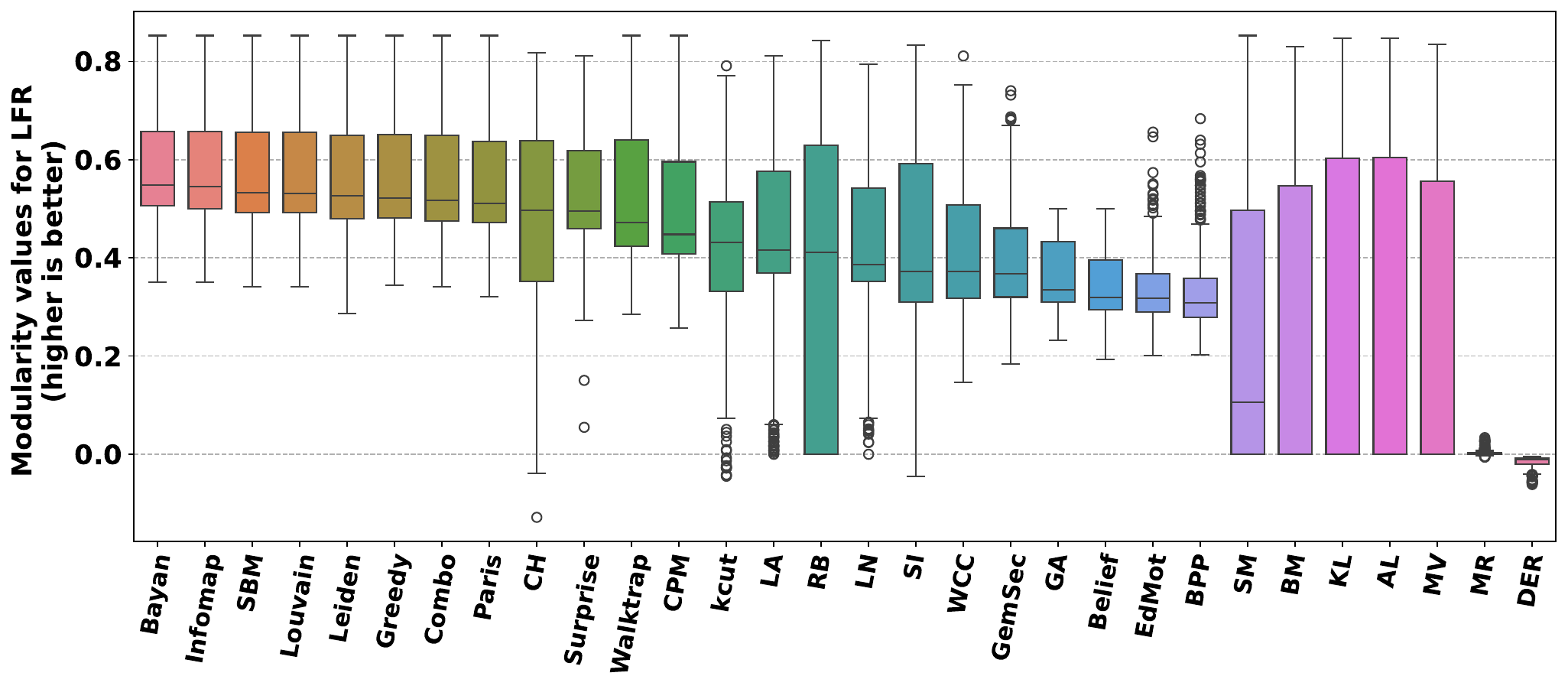}
    \includegraphics[width=0.8\linewidth]{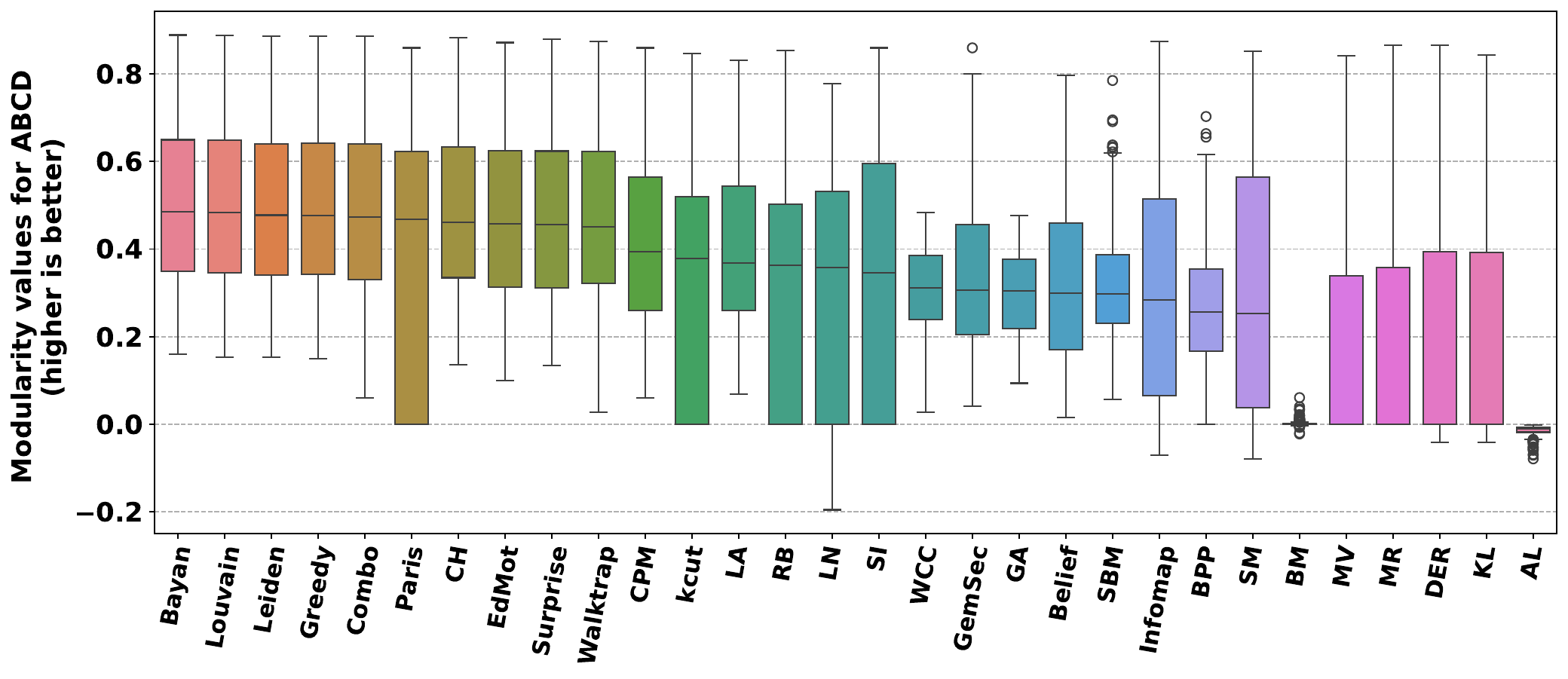}
    \caption{The distribution of modularity values for the partitions produced by each algorithm on the 500 LFR networks (top panel) ad the 500 ABCD networks (bottom panel). The algorithms are sorted from right to left based on having a more desirable median modularity.}
    \label{fig:modularity}
\end{figure}

\begin{figure}
    \centering
    \includegraphics[width=0.8\linewidth]{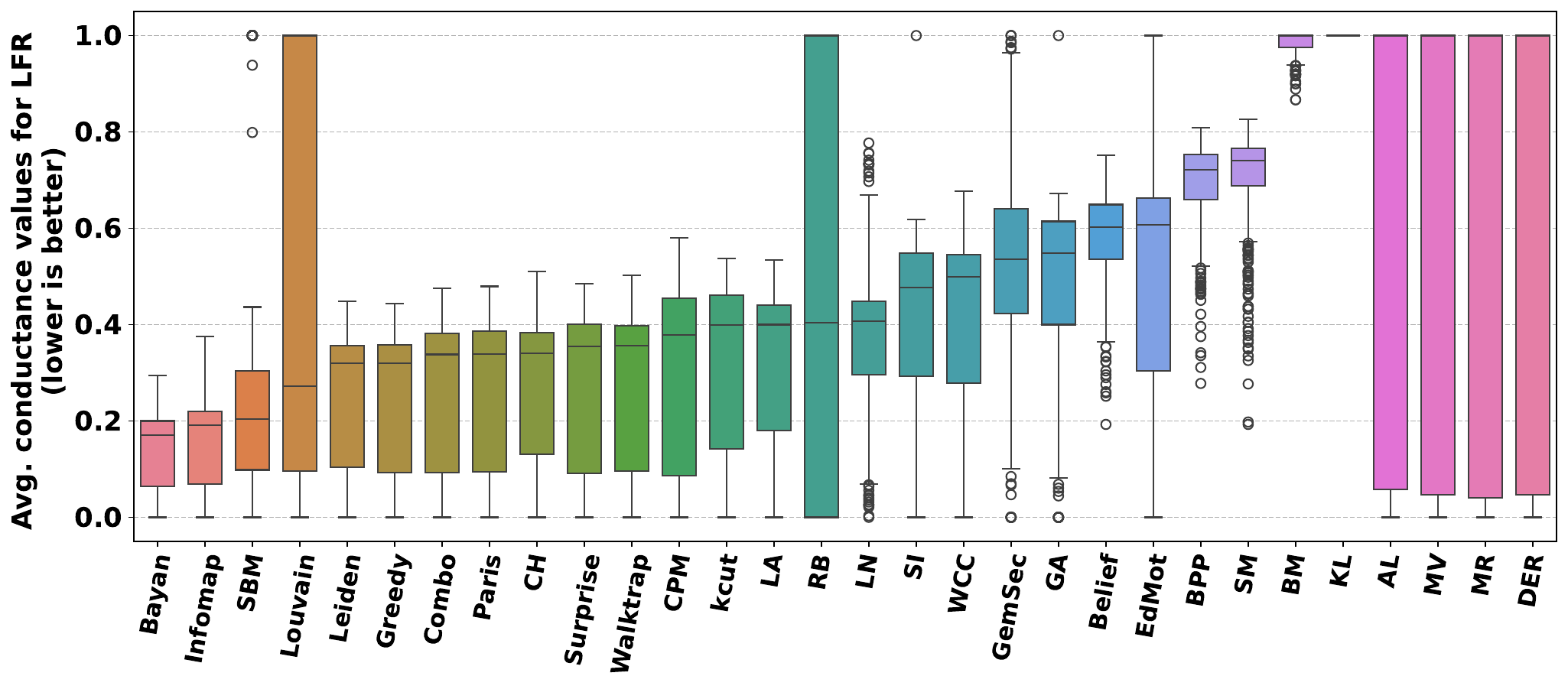}
    \includegraphics[width=0.8\linewidth]{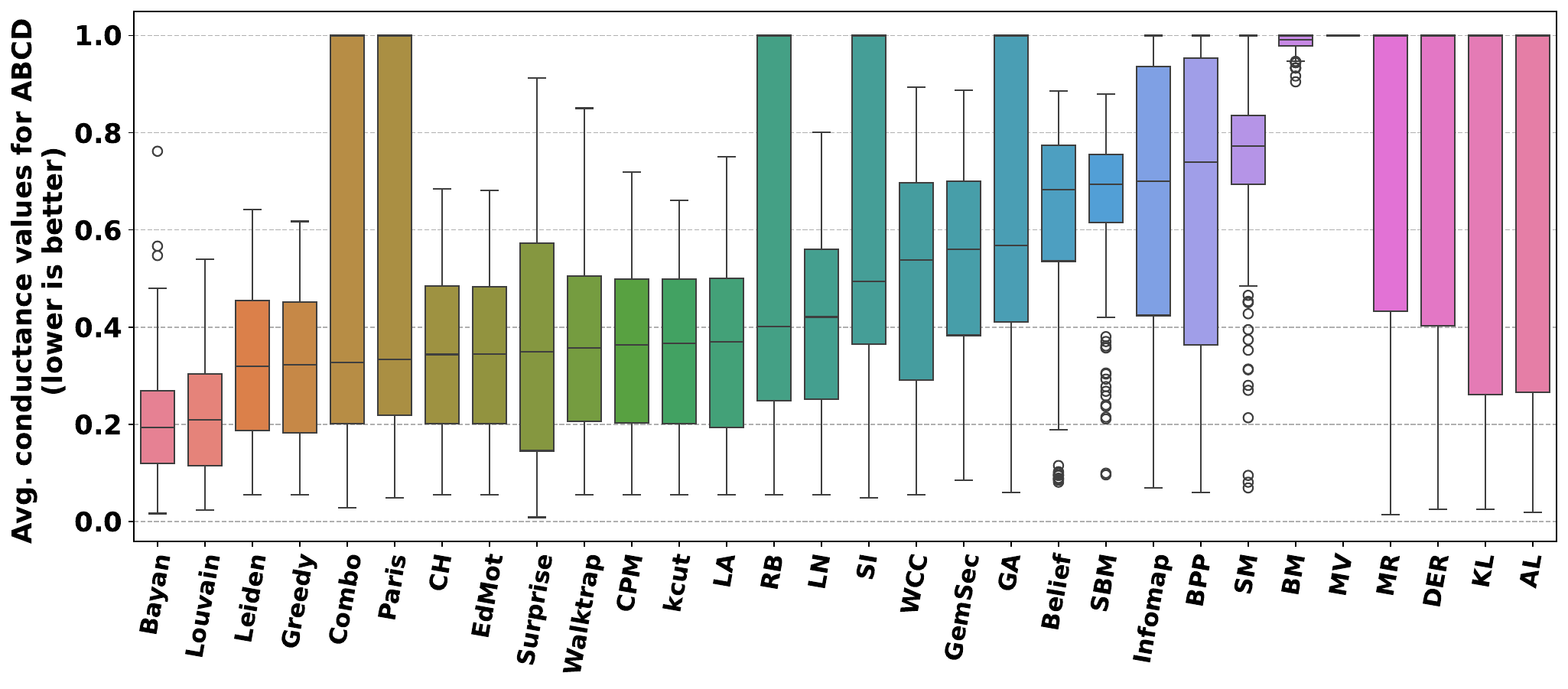}
    \caption{The distribution of average conductance values for the partitions produced by each algorithm on the 500 LFR networks (top panel) ad the 500 ABCD networks (bottom panel). The algorithms are sorted from right to left based on having a more desirable median of average conductance.}
    \label{fig:avg_conductance}
\end{figure}

\begin{figure}
    \centering
    \includegraphics[width=0.8\linewidth]{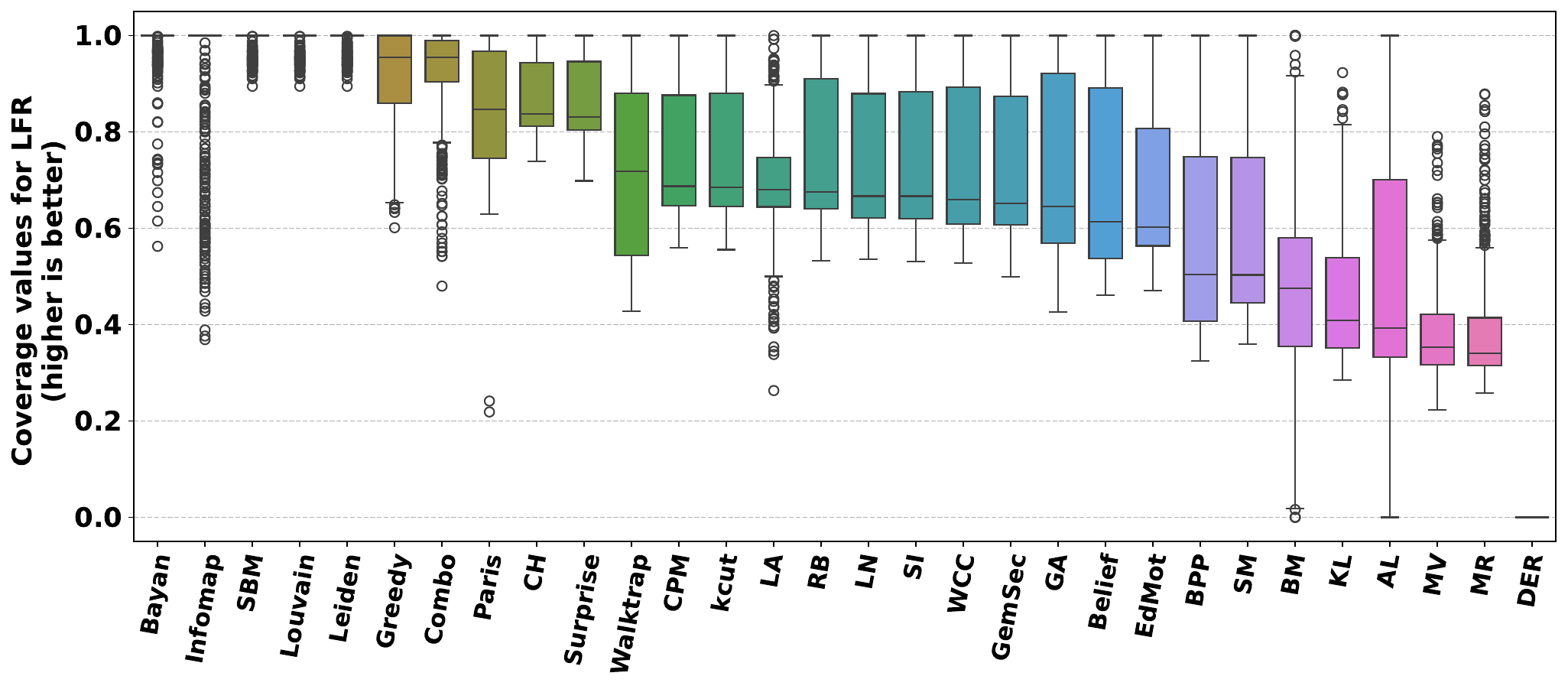}
    \includegraphics[width=0.8\linewidth]{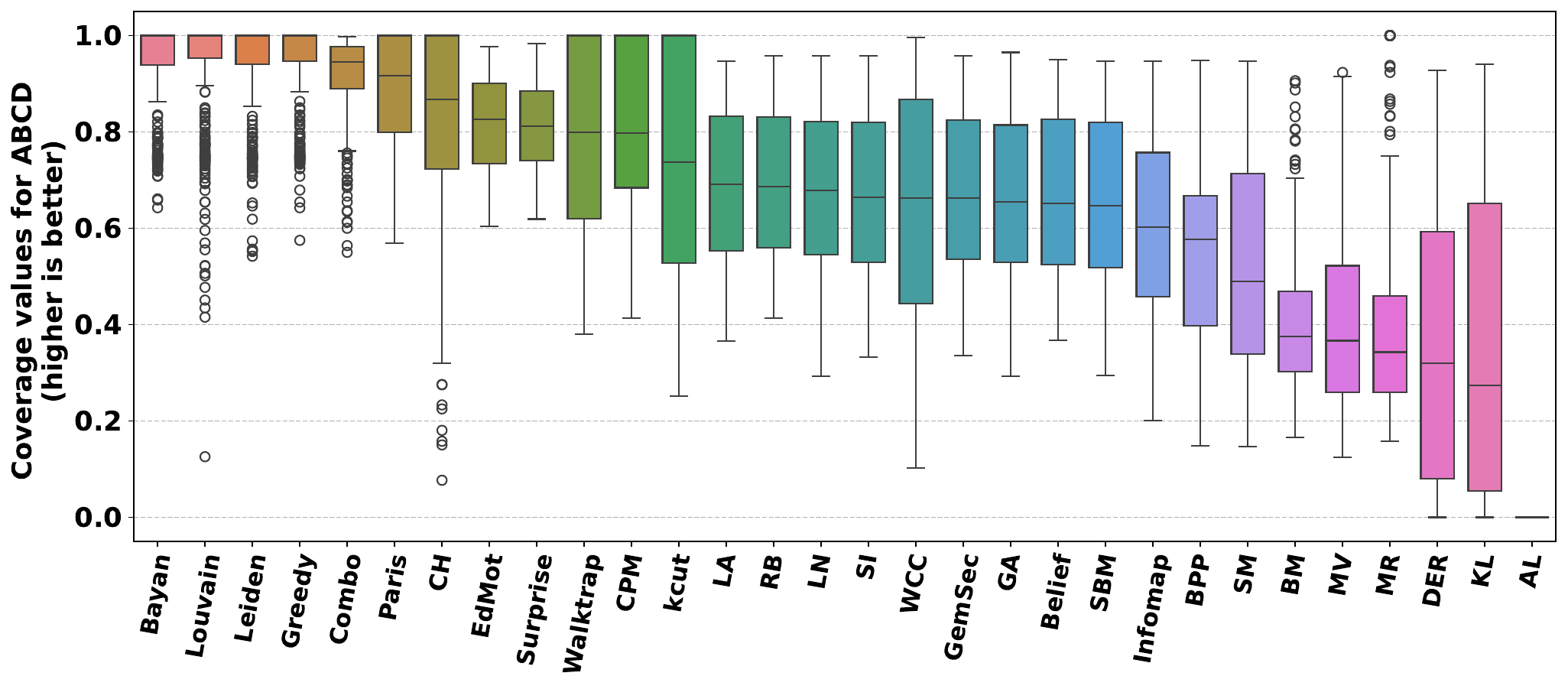}
    \caption{The distribution of coverage values for the partitions produced by each algorithm on the 500 LFR networks (top panel) ad the 500 ABCD networks (bottom panel). The algorithms are sorted from right to left based on having a more desirable median coverage.}
    \label{fig:coverage}
\end{figure}

\begin{figure}
    \centering
    \includegraphics[width=0.8\linewidth]{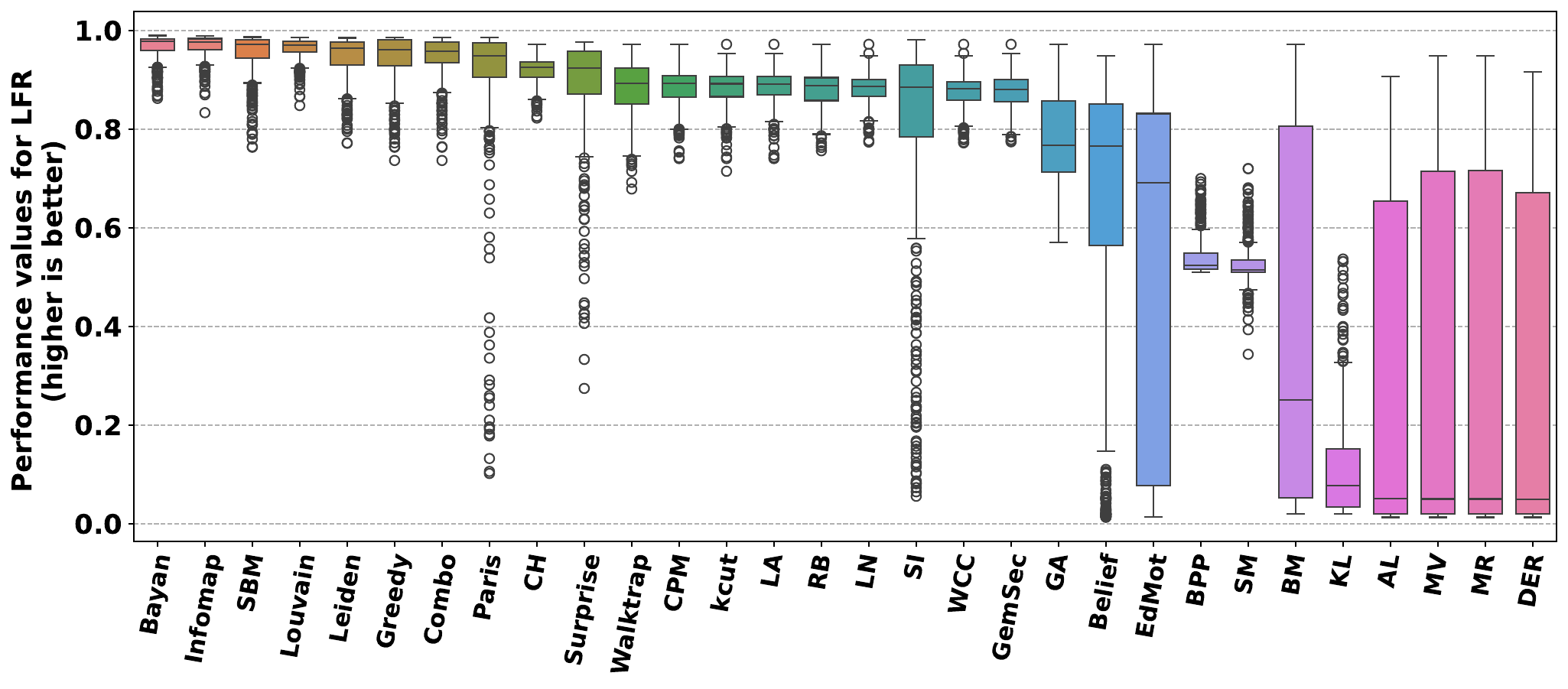}
    \includegraphics[width=0.8\linewidth]{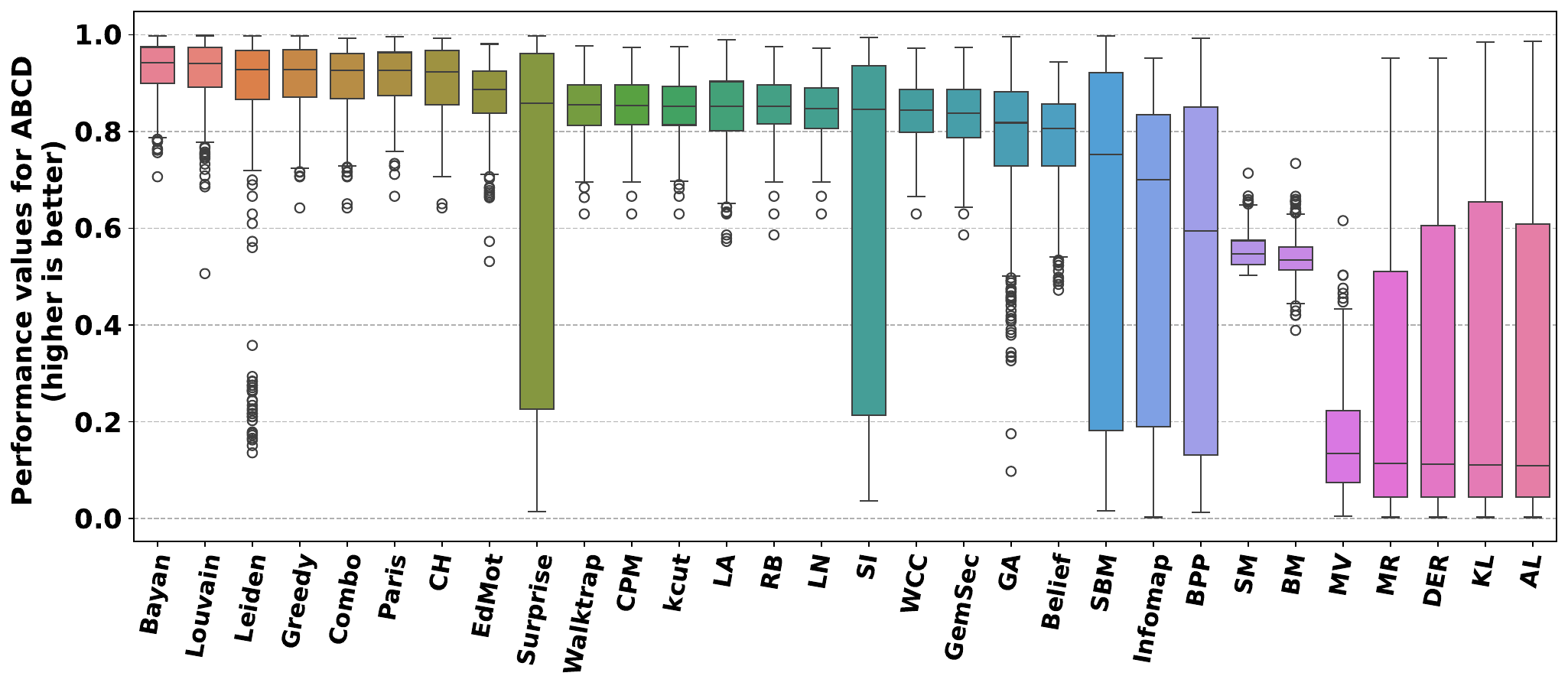}
    \caption{The distribution of performance values for the partitions produced by each algorithm on the 500 LFR networks (top panel) ad the 500 ABCD networks (bottom panel). The algorithms are sorted from right to left based on having a more desirable median performance.}
    \label{fig:performance}
\end{figure}

\begin{figure}
    \centering
    \includegraphics[width=0.6\linewidth]{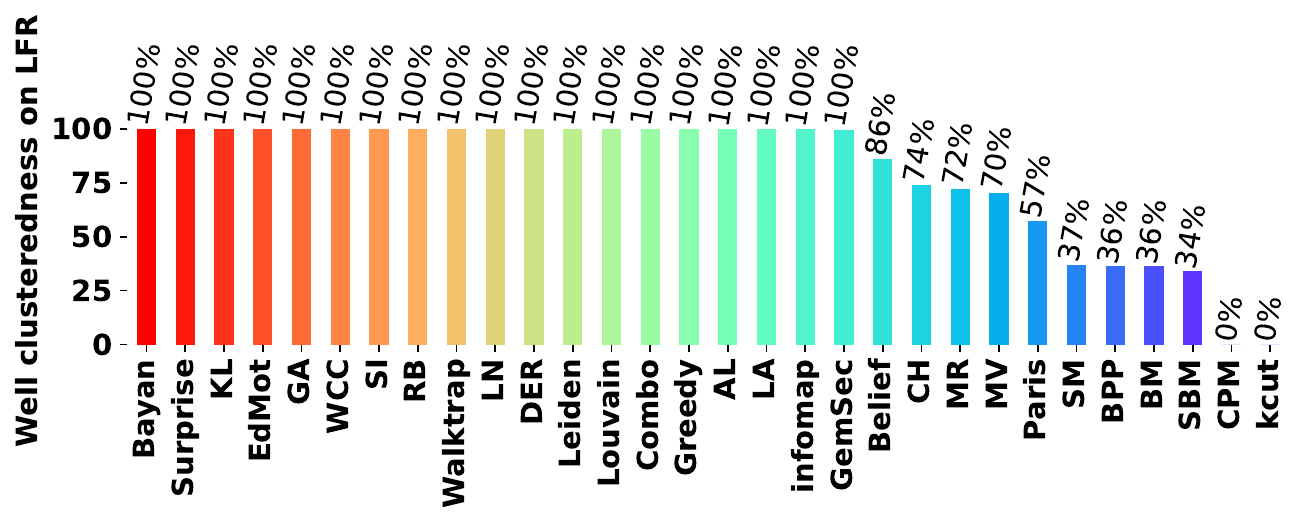}
    \includegraphics[width=0.6\linewidth]{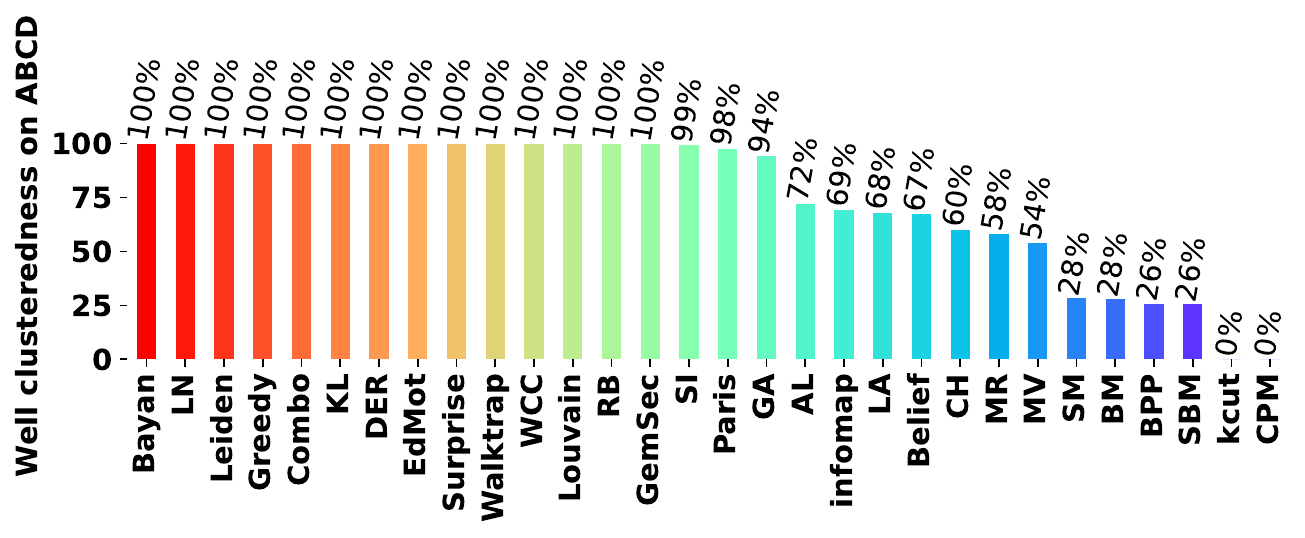}
    \caption{The percentage of LFR (top panel) or ABCD (bottom panel) instances for which each algorithm produces a well clustered graph. Higher values are better. The algorithms are sorted from right to left based on having a more desirable percentage of well clusteredness.}
    \label{fig:well-clusteredness}
\end{figure}

We compute and report the six partition quality measures (defined in Subsection \ref{ss:partition-quality}) for the partitions produced by the $30$ algorithms on the $500$ LFR networks and the $500$ ABCD networks in Figures \ref{fig:description-length}--\ref{fig:well-clusteredness}. The algorithms with the most desirable median value are shown in the left-most positions in Figures \ref{fig:description-length}--\ref{fig:well-clusteredness}.

Figure \ref{fig:description-length} illustrates a box plot for the 500 description length values of each algorithm separately for LFR networks and ABCD networks. Bayan, Louvain, and Leiden have the best (lowest) median values of description length among the 30 algorithms, respectively, for both the LFR and the ABCD instances. Unlike the inferential methods, these three modularity-based algorithms are not designed to minimize the description length. However, they produce partitions with a lower median description length compared to the four inferential algorithms SBM, SM, BPP, and BM. This paradoxical result highlights the importance of a practical comparison of algorithms. Some theoretical motivations seems to be unreliable for choosing algorithms because they may not materialize in practice as illustrated for the inferential algorithms SBM, SM, BPP, and BM in Fig.~\ref{fig:description-length}.

Figure \ref{fig:modularity} illustrates a box plot for the 500 modularity values of each algorithm separately for LFR networks and ABCD networks. As expected Bayan has the best median value of modularity among the 30 algorithms for both the LFR and the ABCD instances. We do not consider this comparison as a fair assessment and only provide it for completeness. Using any single partition quality measure for choosing a CD algorithm is contestable. Moreover, modularity is not a suitable partition quality measure\footnote{There is a distinction between a quality measure (a score function) and an objective (loss) function.} \cite{Miasnikof2020}.

Figure \ref{fig:avg_conductance} illustrates a box plot for the 500 average conductance values of each algorithm separately for LFR networks and ABCD networks. Similarly, the distributions of coverage values are provided in Figure \ref{fig:coverage} and the distributions of performance values are shown in Figure \ref{fig:performance}.

Bayan, Infomap, and SBM have the best median values of average conductance respectively on the LFR networks in Figure \ref{fig:avg_conductance}. The same three algorithms also have the best median values for coverage and performance on LFR networks as shown in Figures \ref{fig:coverage}--\ref{fig:performance}. On ABCD instances, Bayan, Louvain, and Leiden have the best median values of average conductance, respectively. The same three algorithms also have the best median values for coverage and performance on ABCD networks as shown in Figures \ref{fig:coverage}--\ref{fig:performance}.

The results for well clusteredness are reported slightly differently in Figure \ref{fig:well-clusteredness}. Given that well clusteredness is a binary value, we report for each algorithm the percentage of LFR graphs that are well clustered and separately report the similar percentage for ABCD graphs in Figure \ref{fig:well-clusteredness}. The majority of the 30 algorithms including Bayan produce well clustered partitions on all LFR networks. For ABCD networks, there are 14 algorithms including Bayan that produce well clustered partitions on all ABCD instances. For both LFR and ABCD networks, the four inferential algorithms, SBM, SM, BPP, and BM, produce partitions that are not well clustered, more often than not. The two algorithms CPM and kcut never produce well clustered partitions on either LFR or ABCD networks.

In this section, we compared the 30 algorithms based on partition quality measures which do not depend on the planted (ground-truth) partitions. Next, we focus on how the algorithms compare from a run time perspective.

\section{Run Time Comparisons}\label{s:time}

This section provides the last of the three comparisons for the 30 CD algorithms which assesses them based on their run times. We also include two alternative exact modularity maximization methods in Subsection \ref{ss:exact}. These run time analyses are conducted in Python 3.9 using a notebook computer with an Intel Core i7-11800H @ 2.30GHz CPU and 64 GBs of RAM running Windows 10.

\subsection{Comparison of 30 algorithms based on empirical run time}

\begin{figure}
    \centering
    \includegraphics[width=0.85\linewidth]{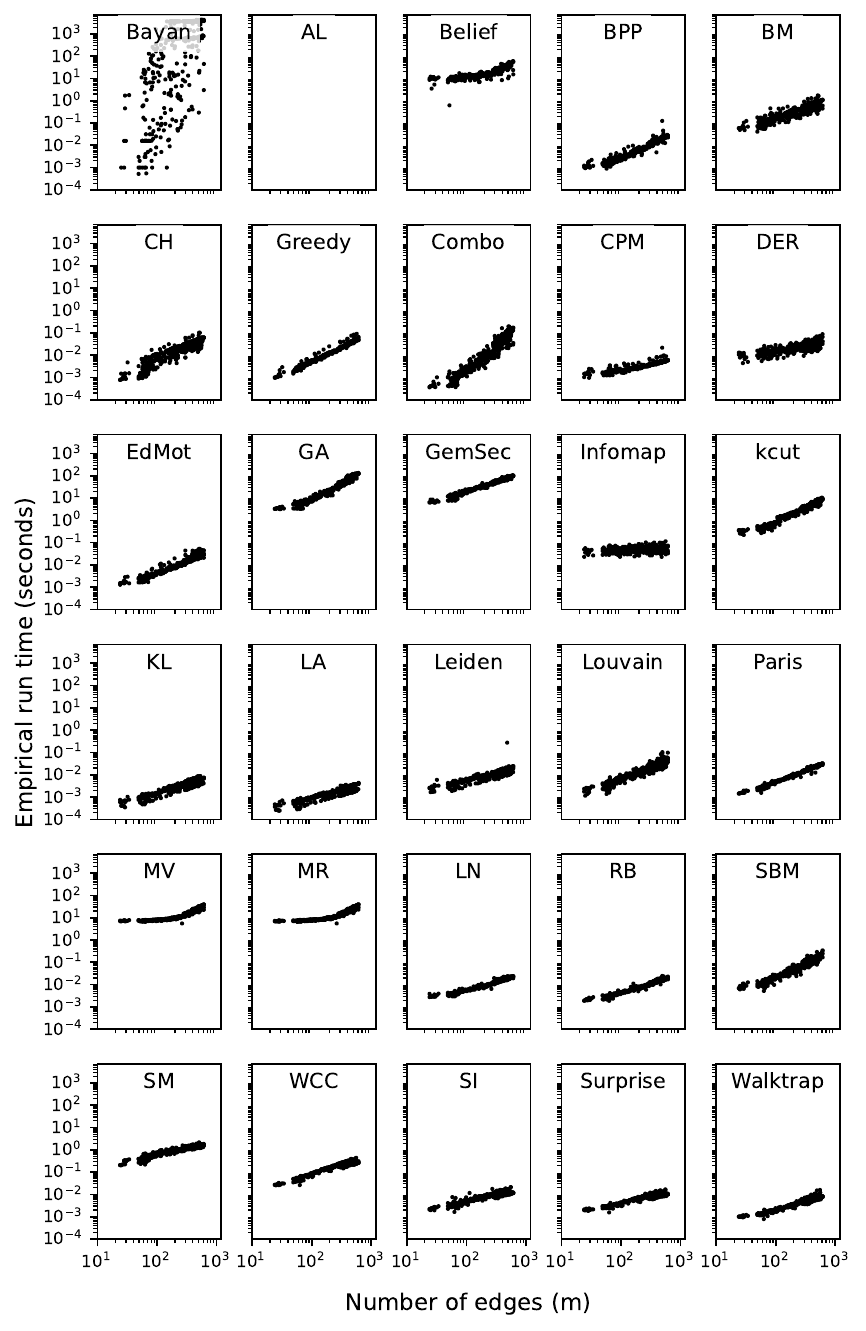}
    \caption{The run time of 30 algorithms on 500 LFR networks. Both axes are scaled logarithmically (base 10).}
    \label{fig:Time_LFR}
\end{figure}

In this subsection, we compare 30 algorithms, including Bayan, based on their empirical run time on the same 500 LFR and 500 ABCD instances. Figure~\ref{fig:Time_LFR} shows scatter plots of the run time (y-axis) for each the 30 algorithms on LFR networks, based on the number of edges (x-axis). The AL algorithm is the fastest algorithm whose run times are all in the order of $1e-5$ seconds and are therefore not visible in the plot. The run times of Bayan have the largest variation ranging mostly in the orders of $1e-3$ to $1e+3$, with larger networks typically taking longer time. For most algorithms except Infomap, we observe a clear increase in run time as the number of edges goes up. The algorithms Bayan, GA, GemSec, and Belief, are the slowest algorithms on the LFR networks. There are many instances for which Bayan has the longest run time. 

\begin{figure}
    \centering
    \includegraphics[width=0.85\linewidth]{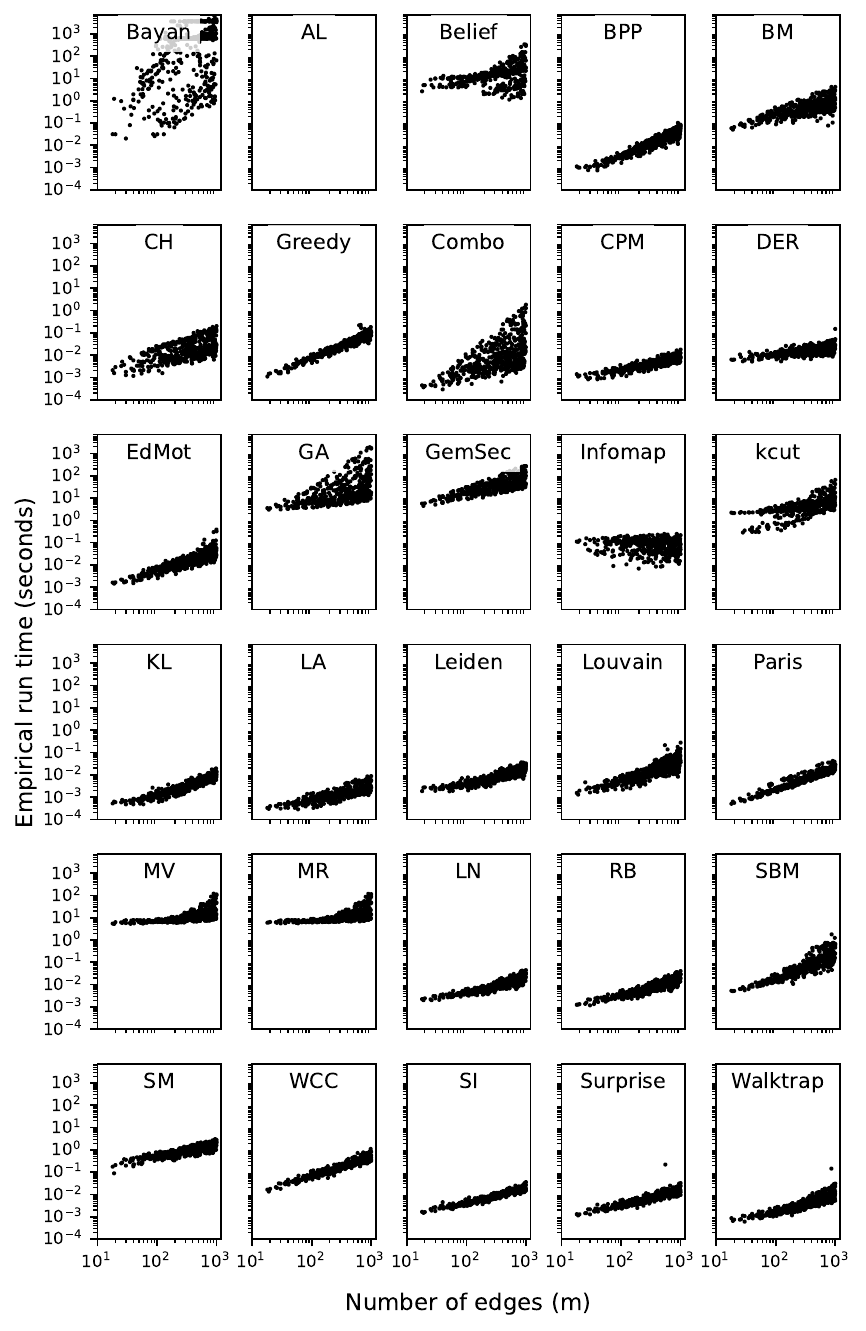}
    \caption{The run time of 30 algorithms on 500 ABCD networks. Both axes are scaled logarithmically (base 10).}
    \label{fig:Time_ABCD}
\end{figure}

Figure~\ref{fig:Time_ABCD} illustrates the run times on ABCD networks for each the 30 algorithms in a scatter plot. The patterns are quite similar to those observed in Figure~\ref{fig:Time_LFR}. The AL algorithm is again observed to be the fastest algorithm with run times around $1e-5$ seconds that are not visible in the scatter plot. The run times of Bayan have the largest variation ranging in the orders of $1e-2$ to $1e+4$ for the ABCD instances (which are on average large networks than our LFR instances). Except Infomap, we observe a clear increase in run time as the number of edges goes up. The four algorithms Bayan, GA, GemSec, and Belief, are the slowest algorithms on the ABCD networks as well. There is difference of several orders of magnitude between the run times of some of these CD algorithms on the same network instances.

Next, we focus on the efficiency of Bayan and two alternative exact approaches for obtaining maximum-modularity partitions.

\subsection{Comparison of three approaches for exact modularity maximization based on time and optimality}
\label{ss:exact}
In this subsection, we compare Bayan with two alternative approaches for exact modularity maximization based on run time and highest modularity found within a maximum limited time. 

The first alternative approach is using the \textit{community\_optimal\_modularity} function\footnote{\href{https://igraph.org/python/doc/api/igraph.Graph.html\#community_optimal_modularity}{https://igraph.org/python/doc/api/igraph.Graph.html}} from the \textit{igraph} library \cite{igraph2006} (IG for short)\footnote{The IG algorithm has not been explicitly proposed as a method for CD, but as a complementary tool in the igraph library \cite{igraph2006}.}. IG calls the open-source solver, GNU Linear Programming Kit (GLPK), to solve an IP formulation \cite{brandes2007modularity} of the modularity maximization problem and returns an optimal partition when/if the optimization problem is solved.

The second alternative approach is solving the sparse IP formulation of the modularity maximization problem \cite{dinh_toward_2015} from Eq.\ \eqref{eq1} using the commercial solver Gurobi \cite{gurobi}. Recent performance assessments from March 2024 show Gurobi to be one of the fastest mathematical solvers for IP models \cite{Miltenberger2023}. This sets a high performance baseline to compare Bayan against. Out-of-the-box Gurobi IP has been used in \cite{aref2023suboptimality} to solve Eq.\ \eqref{eq1} and was shown to be feasible for networks with up to 2812 edges, given relatively long computation time.

To compare IG, Gurobi IP, and Bayan, we consider 94 structurally diverse networks including real networks from different contexts and random graphs. These 94 instances includes 74 real networks with no more than {8405 edges} as well as 10 Erd\H{o}s-R\'{e}nyi graphs and 10 Barab\'{a}si-Albert graphs\footnote{These 94 networks are loaded as simple undirected unweighted graphs. Additional details and information on accessing these 94 networks are provided in the SI.}. 
For each of these 94 instances, we use Bayan, Gurobi IP, and IG with a time limit of 4 hours and compare their empirical run times and objective function values (highest modularity found at termination). 

There are ten instances for which none of the three methods finds a guaranteed optimal partition within the time limit of 4 hours (see Table \ref{tab:tl} in the SM). For these ten instances, Bayan returns a feasible partition with a higher modularity ($49.4\%$ higher on average) compared to the feasible partition from Gurobi IP at the time limit\footnote{Difference in modularity values are reported not as a CD score/quality function, but to compare the objective function values of Bayan and Gurobi IP on the challenging instances.}.

Among the remaining 84 instances, Gurobi IP and IG fail to converge within the time limit respectively on six and 23 others instances respectively. However, all those 84 instances are solved to global optimality by Bayan (detailed results are provided in Tables S4, S5, and S6 in the SM). The six networks \textit{adjnoun}, \textit{london\_transport}, \textit{copenhagen}, \textit{macaques}, 
\textit{malaria\_genes}, and \textit{celegans\_2019}\footnote{See the network repository \href{https://networks.skewed.de/}
{Netzschleuder} for more information on the networks.}
cannot be solved by Gurobi IP or IG within the time limit. To our knowledge, the guaranteed maximum-modularity partitions for these six networks have not been previously determined \cite{aloise_column_2010,miyauchi2013computing,cafieri2014reformulation,dinh_toward_2015,dinh_network_2015,sobolevsky_optimality_2017,aref2023suboptimality}.

On these 84 instances, the average run times of Bayan, Gurobi IP, and, IG are 
295.08, 1110.60, and 4588.52 seconds respectively. This implies that on average, Bayan is 815.52 seconds per instance (or over 3.7 times) faster than the Gurobi IP, and 4289.15 seconds per instance (or over 15.5 times) faster than the IG on these 84 instances. Both of these run time comparisons are conservative because Bayan solves all these 84 instances to global optimality while Gurobi IP and IG fail to terminate within the time limit for six and 23 networks respectively. On all these failed cases, we have conservatively considered 4 hours as the run time for Gurobi IP and IG while the actual run time may be substantially higher. Therefore, the actual performance comparison would be even more favourable to Bayan. The complete results on run times and modularity values of the three exact methods are provided in the SM (see Tables S3-S6). 

Overall, these results indicate that Bayan is considerably faster than IG and Gurobi IP for most of the networks considered. The run time advantage of Bayan makes a practical difference on the real network instances and especially the six challenging instances noted earlier which are solvable by Bayan but not solvable by IG or Gurobi IP (or any other exact method to the best of our knowledge).

\section{Discussion and Future Directions}\label{s:discuss}

%two questions
There are two fundamental questions to be answered regarding the Bayan algorithm:\\ (1) How suitable is Bayan for maximizing modularity?\\ (2) How suitable is Bayan for community detection? 

For the first question, Fig. \ref{fig:modularity} provides some answers. For more conclusive answers about global maximization of modularity, one can also refer to another study \cite{aref2023analyzing}, where we have compared ten modularity-based methods based on the extent to which they globally maximized modularity. This included Bayan, eight modularity-based heuristics and a graph neural network algorithm \cite{sobolevsky2022gnn}. Using real and synthetic networks with fairly modular structure, we showed the following: At the cost of longer computation time, Bayan has the advantage of reaching global optimality at higher rates and returning partitions more similar to the globally optimal partitions, compared to the other 9 modularity maximization algorithms. Bayan was followed by the Combo algorithm as the heuristic method demonstrating the most promising performance in returning partitions that were optimal or were similar to an optimal partition.

The results provided in Sections \ref{s:retrieval}--\ref{s:quality} allow us to answer the second question from a practical standpoint. The results in Subsections \ref{ss:results_lfr}--\ref{ss:results_abcd} demonstrate the comparative suitability of Bayan based on retrieving planted partitions of LFR and ABCD benchmarks. Bayan was observed to be among the top algorithms in  retrieving planted partitions across different mixing parameters. The results in Sections \ref{ss:results_real} show that, for networks where node labels are aligned with the structure, Bayan retrieves partitions that are similar with node labels. Overall, the results in Section \ref{s:retrieval} demonstrate the suitability of Bayan as a community detection method for networks with up to 3000 edges. Our results in Section \ref{s:quality} show Bayan's competitive advantages in comparison with 29 other methods based on five partition quality measures other than modularity: description length, average conductance, coverage, performance, and well clusteredness. In what follows, we discuss five key insights from our results.

\paragraph{Global modularity maximization retrieve planted partitions at rates higher than other existing algorithms}
The results provided in Section \ref{s:retrieval} indicate the practical relevance of maximum-modularity partitions in comparison with 29 other alternatives for CD. Despite the well-document theoretical flaws of modularity \cite{peixoto_2023}, the high accuracy and stability of globally maximum-modularity partitions (shown in this study) are not challenged by other existing methods for community detection. 

\paragraph{In modularity optimization,
guaranteed closeness to optimality matters.} The results from Figures \ref{fig:RankingLFR} -- \ref{fig:RankingABCD} and Figures \ref{fig:description-length} -- \ref{fig:well-clusteredness} offer additional insights when we focus on the performance of the nine algorithms that attempt to maximize modularity. The comparative performance of nine modularity-based algorithms (Bayan, greedy, Louvain, LN, Combo, Belief, Paris, Leiden, and EdMot) shows the practical benefits of global optimization in the context of using modularity for CD. The practical benefit of global optimization is in both achieving better performance in retrieving planted communities and achieving better partition quality measures (other than modularity). The key lesson learned is that in the context of community detection through modularity optimization, guaranteed closeness to optimality matters. This crucial, yet often overlooked \cite{Kosowski2020}, lesson was foreshadowed in \cite[pp. 012811-5]{sobolevsky2014general} which reads ``to stress the importance of looking for even the minor gains in the modularity score, [...] relatively small changes in this partition quality function can be reflected by macroscopic variation of the communities involved".

\paragraph{Exact modularity maximization is within reach for small networks, but requires substantially longer computation.}
The scatter plots provided in Figures \ref{fig:Time_LFR}--\ref{fig:Time_ABCD} demonstrate the substantial differences between the run times of algorithms. As expected and being an exact method, Bayan takes considerably longer than most heuristic methods. However, it is, on average, over 15.5 times faster than using the open-source IP solver, GLPK, and over 3.7 times faster than using the commercial solver, Gurobi, for modularity maximization. Also, we reiterate that Bayan (as well as any attempt at solving an NP-hard problem exactly) does not scale to large instances and suffers from asymptotic worst-case running times that are far from ideal. The NP-hardness of the problem implies that efficient running times of scalable heuristics (like Leiden) cannot be beaten by developing an exact method for solving it; scaling to large networks has not been a purpose of this study but the topic of a tempting and recurring questions. Having said that, exact and approximate optimization has provided more accurate and reliable solutions to increasingly larger instances of many similar NP-hard graph partitioning problems \cite{aref_modeling_2020,aref2021identifying,bonchi2022correlation,belyi2023subnetwork,sukeda2023study} and modularity maximization is not an exception. Using decades-old heuristics (including methods that were groundbreaking for their time) is not justified for small networks and come at the cost of lower performance.

\paragraph{Bayan's mathematically rigorous operationalization of modularity maximization is a reliable choice for CD in small networks with up to 3000 edges.} While it is tempting to interpret our results as suggesting modularity is a better objective function than the objective functions of other optimization-based algorithms considered, we refrain from drawing such a conclusion because our comparative results of algorithms in Figures \ref{fig:RankingLFR}--\ref{fig:RankingABCD} and Figures \ref{fig:description-length} -- \ref{fig:well-clusteredness} are confounded by the difference in the objective functions and the difference between mathematically rigorous optimization vs.\ local/greedy/heuristic optimization. We simply recommend using mathematically rigorous optimization in optimization-based tasks for small input. Our results justify the exact optimization approach for modularity where optimality matters \cite{sobolevsky2014general} and sub-optimality has a disproportionate cost \cite{aref2023suboptimality}. Future research may investigate the potential advantages of using global optimization and guaranteed approximations for other objective functions (e.g., Markov stability \cite{pygenstability_2023}, description length \cite{sbm_2014}, modularity density \cite{sato_enhanced_2019}, surprise \cite{surprise_2015,marchese2022detecting}, or a new objective function inspired by the map equation \cite{rosvall_2008}) for CD. To the best of our knowledge, exact optimization of these alternative objectives are underexplored, and therefore each has the potential to outperform maximum-modularity partitions (and in turn outperform most other algorithms considered in this study) in retrieval tests and partition quality measures. We hope that our publicly available data and algorithm facilitate these prospective advances in the field.
As a \textit{specialized algorithm} \cite{peel2017ground} for accurate CD in small networks, Bayan maximizes modularity in %unweighted and weighted 
networks with up to 3000 edges (in their largest connected component) and approximates maximum modularity in slightly larger instances on ordinary computers. Prospective advances in IP solvers (like Gurobi which is used in Bayan and is actively improved) pushes this limit further. %Bayan offers a unique method for detecting communities with a guarantee of optimality. 
Exact maximization of modularity for larger networks may require a substantially longer time or high performance computing resources. For slightly larger networks on an ordinary computer, one may run Bayan with a desired optimality gap tolerance (e.g. $0.1$) or a desired time limit so that it returns a partition with a guaranteed proximity to the maximum modularity. 

\paragraph{Bayan contributes computational capabilities of both methodological and empirical relevance.} On the methodological side, the contributions of Bayan is threefold. (1) Bayan offers a principled way of quantifying the optimality of current and future modularity-based heuristics for CD. (2) Bayan demonstrates the advantages of a mathematically rigorous implementation of a long-lasting yet paradoxically underexplored optimization idea. (3) Bayan reaffirms that the practical relevance of modularity as an objective function, despite its theoretical flaws \cite{peixoto_2023}, is yet to be challenged by a practical method that is more accurate and stable. On the empirical side, Bayan tackles a practical task in network science for small input, thereby improving upon open-access computational capabilities for a narrow but relevant set of tasks in networked systems.

\subsection*{Data and materials availability} 
A Python implementation of the Bayan algorithm (\textit{bayanpy}) is publicly available through the \href{https://pypi.org/project/bayanpy/}{package installer for Python (pip)}. All network data used in the experiments and evaluations of the Bayan algorithm are available in a \textit{FigShare} data repository \cite{Aref2023figshare}.

\subsection*{CRediT Author Contribution Statement} 
Conceptualization (SA); data curation (SA, HC); formal analysis (SA, MM); funding acquisition (SA); investigation (SA, MM); methodology (SA, MM); project administration (SA); resources (SA, MM, HC); software (SA, MM, HC); supervision (SA, MM); validation (SA, MM, HC); visualization (MM); writing - original draft preparation (SA, MM); writing - review \& editing (SA, MM, HC).

% \subsection*{Declaration of Competing Interest} 
% Authors do not have any conflicts of interest to declare.

\subsection*{Acknowledgements} 
Authors acknowledge Sanchaai Mathiyarasan for technical assistance and Giulio Rossetti, R\'emy Cazabet, Pawel Pralat, and Tiago P. Peixoto for helpful discussions. This study has been supported by the Data Sciences Institute at the University of Toronto.

%\appendix

\clearpage

\section*{Appendix: Detailed results of 30 community detection algorithms on 500 LFR graphs}

%\subsection{Average AMI results on 500 LFR graphs}

Our retrieval tests on LFR benchmarks produce 100 AMI values for each of the 30 algorithms in each of the 5 LFR experiment settings. The average AMIs are reported in Table I.

\begin{table}[htbp!]
\label{tab:lfr}
\small
\centering
\caption{Average AMIs between the partitions of each algorithm and the ground-truth partitions of LFR graphs. Each column shows average AMI for 100 LFR graphs generated with a specific value for the parameter $\mu$.}
\begin{tabular}{lccccc}
\hline
Algorithm \textbackslash \quad $\mu$ & 0.01 & 0.1 & 0.3 & 0.5 & 0.7 \\ \hline
Bayan & 0.914977 & 0.865469 & 0.222458 & 0.120004 & 0.048301 \\
Belief & 0.807955 & 0.779013 & 0.15946 & 0.048613 & 0.021865 \\
BPP & 0.883869 & 0.682873 & 5.39E-17 & 5.3E-17 & 5.3E-17 \\
BM & 0.884644 & 0.680429 & 5.39E-17 & 5.3E-17 & 5.3E-17 \\
CH & 0.900439 & 0.660127 & 0.114033 & 0.043228 & 0.015612 \\
Combo & 0.911275 & 0.81159 & 0.20871 & 0.10669 & 0.043502 \\
CPM & 5.57E-14 & -9.3E-15 & -8.4E-15 & -2.4E-14 & -1.7E-14 \\
DER & 0.566681 & 0.511116 & 0.090219 & 0.048349 & 0.023683 \\
EdMot & 0.812163 & 0.799894 & 0.215178 & 0.116701 & 0.046678 \\
GA & 0.729344 & 0.6357 & 0.178047 & 0.098336 & 0.03072 \\
GemSec & 0.471457 & 0.464521 & 0.159129 & 0.099382 & 0.036814 \\
Greedy & 0.906994 & 0.803937 & 0.194709 & 0.102737 & 0.046088 \\
Infomap & 0.892984 & 0.7848 & 0.217829 & 0.124187 & 0.045835 \\
kcut & 0.026286 & 0.004739 & -0.00038 & -0.00243 & -0.00216 \\
KL & 0.507826 & 0.457056 & 0.087676 & 0.036781 & 0.016813 \\
AL & 0.747412 & 0.68016 & 0.209216 & 0.111949 & 0.033243 \\
LA & 0.839582 & 0.716382 & 0.204512 & 0.095657 & 0.03089 \\
Leiden & 0.869695 & 0.803111 & 0.215096 & 0.111227 & 0.042612 \\
Louvain & 0.880473 & 0.798383 & 0.202008 & 0.108993 & 0.038464 \\
MV & 0.224506 & 0.693942 & 0.178752 & 0.104611 & 0.038645 \\
MR & 0.400319 & 0.61916 & 0.207398 & 0.116707 & 0.04327 \\
Paris & 0.992753 & 0.457702 & 0.052421 & 0.03215 & 0.017041 \\
LN & 0.900007 & 0.805143 & 0.213839 & 0.110176 & 0.043872 \\
RB & 0.851303 & 0.808343 & 0.233574 & 0.12645 & 0.045674 \\
SBM & 0.785009 & 0.60195 & 5.39E-17 & 5.3E-17 & 5.3E-17 \\
SM & 0.78293 & 0.637965 & 5.39E-17 & 5.3E-17 & 5.3E-17 \\
SI & 0.565777 & 0.465077 & 0.193266 & 0.102577 & 0.038334 \\
Surprise & 0.015216 & 0.014727 & 0.003753 & 0.001559 & -0.00053 \\
Walktrap & 0.948559 & 0.903521 & 0.219744 & 0.117848 & 0.035803 \\
WCC & 0.417899 & 0.452965 & 0.162383 & 0.107579 & 0.038922 \\ \hline
\end{tabular}
\end{table}

%\FloatBarrier

\clearpage

\section*{Appendix: Detailed AMI results of 30 community detection algorithms on 500 ABCD graphs}

Our retrieval tests on ABCD benchmarks produce 100 AMI values for each of the 30 algorithms in each of the 5 ABCD experiment settings. The average AMIs are reported in Table II.

%\subsection{Average AMI results on 500 ABCD graphs}

\begin{table}[htbp!]
\label{tab:abcd}
\small
\centering
\caption{Average AMIs between the partitions of each algorithm and the ground-truth partitions of ABCD graphs. Each column shows the average AMI for 100 ABCD graphs generated with a specific value for the parameter $\xi$.}
\begin{tabular}{lccccc}
\hline
Algorithm \textbackslash \ $\xi$ & 0.1 & 0.3 & 0.5 & 0.7 & 0.9 \\ \hline
Bayan & 0.904262 & 0.768595 & 0.516145 & 0.172341 & 0.039422 \\
Belief & 0.872381 & 0.68601 & 0.402638 & 0.044745 & 0.006225 \\
BPP & 0.757256 & 0.476529 & 0.042167 & 0 & 0 \\
BM & 0.805145 & 0.505006 & 0.040564 & 6.63E-06 & 0 \\
CH & 0.887096 & 0.64501 & 0.09825 & 0.020215 & 0.006525 \\
Combo & 0.906193 & 0.763176 & 0.497164 & 0.170715 & 0.033017 \\
CPM & -1.21E-14 & 1.16E-14 & 2.72E-14 & -5.62E-15 & -3.96E-14 \\
DER & 0.529986 & 0.405244 & 0.220585 & 0.069547 & 0.01587 \\
EdMot & 0.845659 & 0.684886 & 0.450476 & 0.163103 & 0.044221 \\
GA & 0.670259 & 0.416827 & 0.217894 & 0.139202 & 0.035661 \\
GemSec & 0.638734 & 0.480496 & 0.26845 & 0.108372 & 0.029334 \\
Greedy & 0.878954 & 0.675435 & 0.37166 & 0.135652 & 0.025767 \\
Infomap & 0.830611 & 0.670542 & 0.378866 & 0.102234 & 0.020956 \\
kcut & -0.0019 & 0.000836 & 0.001912 & -0.00024 & 0.002985 \\
KL & 0.452558 & 0.384772 & 0.244365 & 0.072133 & 0.014224 \\
AL & 0.74617 & 0.628092 & 0.233951 & 0.069981 & 0.011225 \\
LA & 0.764068 & 0.573115 & 0.167497 & 0.05184 & 0.010451 \\
Leiden & 0.901538 & 0.744484 & 0.492217 & 0.161479 & 0.035678 \\
Louvain & 0.877696 & 0.713213 & 0.453911 & 0.147747 & 0.039236 \\
MV & 0.870493 & 0.705155 & 0.408535 & 0.14907 & 0.040766 \\
MR & 0.744999 & 0.615564 & 0.424862 & 0.181526 & 0.040006 \\
Paris & 0.799557 & 0.511702 & 0.265896 & 0.108647 & 0.03216 \\
LN & 0.902227 & 0.735016 & 0.482786 & 0.158218 & 0.043241 \\
RB & 0.897224 & 0.73453 & 0.500524 & 0.149088 & 0.034994 \\
SBM & 0.72774 & 0.46625 & 0.077345 & -0.0005 & -0.00352 \\
SM & 0.708577 & 0.44547 & 0.040516 & -0.00614 & 0.005209 \\
SI & 0.598734 & 0.478368 & 0.33308 & 0.150461 & 0.04024 \\
Surprise & 0.790037 & 0.645862 & 0.42343 & 0.165148 & 0.049853 \\
Walktrap & 0.899926 & 0.745977 & 0.475442 & 0.170127 & 0.032478 \\
WCC & 0.524841 & 0.474244 & 0.346609 & 0.1596 & 0.041969 \\ \hline 
\end{tabular}
\end{table}

\FloatBarrier

%\bibliographystyle{ieeetr}
%\bibliography{refs}

\end{document}